%% file: main.tex
\documentclass[sigconf,screen,nonacm]{acmart}

\AtBeginDocument{%
  \providecommand\BibTeX{{%
    \normalfont B\kern-0.5em{\scshape i\kern-0.25em b}\kern-0.8em\TeX}}}

\usepackage{enumitem}
\usepackage{quoting}
\usepackage[most]{tcolorbox}

\usepackage{listings}
\usepackage{xcolor}

\lstdefinestyle{mypython}{
  language=Python,
  basicstyle=\ttfamily\footnotesize,
  keywordstyle=\color{blue}\bfseries,
  stringstyle=\color{red!70!black},
  commentstyle=\color{green!50!black}\itshape,
  numberstyle=\tiny\color{gray},
  numbers=right,
  stepnumber=1,
  frame=single,
  breaklines=true,
  showstringspaces=false,
  tabsize=4,
  xleftmargin=5pt,
  xrightmargin=5pt,
  framexleftmargin=5pt,
  framexrightmargin=5pt,
  aboveskip=2pt,
  belowskip=2pt,
  lineskip=-1pt
}

\usepackage{framed}

\usepackage{parskip}
\usepackage{graphicx}
\usepackage{subcaption}
\usepackage{tabularx, array, booktabs}

\begin{document}


\title[Beyond the Traceback: Using LLMs for Adaptive Explanations of Programming Errors]{Beyond the Traceback: Using LLMs for Adaptive Explanations of Programming Errors}

\author{Alexandru-Radu Moraru}
\affiliation{%
  \institution{Delft University of Technology}
  \city{Delft}
  \country{The Netherlands}}
\email{alexradumoraru@gmail.com}

\author{Shreyan Biswas}
\affiliation{%
  \institution{Delft University of Technology}
  \city{Delft}
  \country{The Netherlands}}
\email{s.biswas@tudelft.org}

\author{Ujwal Gadiraju}
\affiliation{%
  \institution{Delft University of Technology}
  \city{Delft}
  \country{The Netherlands}}
\email{u.k.gadiraju@tudelft.nl}

\renewcommand{\shortauthors}{Moraru et al.}


\input{tex/0_abstract}

\maketitle
\input{tex/1_Introduction}
\input{tex/2_Background}
\input{tex/3_Experiment_Design}
\input{tex/5_Results}
\input{tex/6_discussion}
\input{tex/7_conclusion}
\section*{Acknowledgements}
The authors of this paper used large language models to rephrase certain sentences and to assist in drafting some parts of the programming work, such as components of the experimental web application interface, data analysis utilities, and scripts for generating figures. All system outputs were carefully reviewed and validated by the authors before usage.

\bibliographystyle{ACM-Reference-Format}
\bibliography{references}
\appendix

\input{appendix/appendix_study_design}
\input{appendix/appendix_a}
\input{appendix/appendix_b}
\input{appendix/appendix_c}

\end{document}

%% file: tex/0_abstract.tex
\begin{abstract}

Programming error messages are critical for software development, yet they remain difficult for novice programmers to interpret. While Large Language Models (LLMs) can rewrite these errors into clearer explanations, it remains unclear whether increased readability improves objective debugging performance or how explanation styles should align with programmer skill. We present a multi-stage crowdsourced study ($N=103$) evaluating skill-targeted, LLM-generated Python error messages. Using a custom proficiency assessment, we categorized participants by skill level and tested standard interpreter messages against two LLM-generated styles: \textit{pragmatic} (action-oriented) and \textit{contingent} (scaffolded explanations). We measured both objective debugging metrics (fix rate, attempts, time-to-fix) and subjective perceptions (readability, cognitive load, tone). Our results show that while LLM-rewritten messages significantly improved subjective evaluations, with pragmatic messages rated as clearer and less cognitively demanding, these perceived gains did not translate into statistically significant improvements in objective debugging performance. This highlights a critical human-AI complementarity gap: explanations that feel better to users do not necessarily make them more effective debuggers. We discuss design implications for adaptive AI feedback systems, arguing that future tools should pivot from static skill-targeted rewriting toward dynamic adjustments based on a user's real-time repair trajectory.

\end{abstract}

%% file: tex/1_Introduction.tex
\section{Introduction}
\label{sec:introduction}

Programming error messages are one of the most frequent forms of feedback that programmers receive from software systems. When a program fails, the compiler or interpreter becomes an immediate explanatory interface: it identifies that something went wrong, points to a location, and offers diagnostic information that the programmer must interpret and act upon. In principle, such messages should support learning and repair. In practice, they often remain difficult to understand, especially for less experienced programmers who may lack the conceptual vocabulary and debugging experience needed to map a diagnostic message to an actionable fix. Even experienced programmers can spend substantial effort interpreting error messages, locating the underlying fault, and deciding how to proceed.

Prior work has repeatedly shown that debugging is time-consuming, frustrating, and cognitively demanding across both educational and professional settings \cite{what-constitutes-debugging, obi2024irritating, debugging-practices}. Poorly designed error messages can slow progress, increase mental effort, and undermine learners' confidence \cite{attrition-cs, drop-out-predictions}. A survey by \citet{compiler-error-messages-considered-unhelpful} identifies several recurring weaknesses in programming error messages, including excessive jargon, insufficient context, poor localization, and limited actionable guidance. These limitations have motivated a long line of work on improving compiler and interpreter diagnostics through clearer wording, better examples, enhanced localization, and pedagogically informed feedback \cite{widjojo2023addressingcompilererrorsstack, code-debugging-llm-generated-explanations, pycee-stack-overflow}.

Large language models (LLMs) introduce a new opportunity for improving programming error messages. Rather than relying only on fixed diagnostic templates, LLMs can rewrite interpreter output using the surrounding code, traceback, and user-facing design goals. In principle, such rewrites could make error messages more readable, less intimidating, and more actionable. However, prior work on scaffolding, cognitive load, and expertise-sensitive instruction suggests that feedback is rarely universally effective: support that helps less experienced learners can become redundant, distracting, or inefficient for more experienced users \cite{wood1976role, sweller2011cognitive, kalyuga2007expertise}. This creates a human computation problem for LLM-generated programming feedback: how should such messages be designed, calibrated, and evaluated with human users? A rewritten explanation that sounds helpful may still fail to improve debugging performance. Similarly, an explanation that is useful for a more experienced programmer may be too terse for a less experienced programmer, while a scaffolded explanation that supports a less experienced programmer may feel verbose or distracting to someone with stronger debugging fluency. Thus, the design of LLM-generated programming feedback requires not only model capability, but also systematic human evaluation across user skill levels and debugging contexts.

In this paper, we study skill-targeted LLM rewrites of Python programming error messages through a multi-stage crowdsourced evaluation. We use the term \textit{skill-targeted} to describe explanation styles designed for different assessed proficiency levels; we do not claim that the system dynamically adapts during interaction. We focus on two explanation styles. \textit{Pragmatic} messages provide concise, action-oriented guidance intended to help users move quickly from diagnosis to repair. \textit{Contingent} messages provide more scaffolded explanatory support, offering additional context about the likely cause of the error and how the programmer might reason about it. Rather than treating LLM-generated explanations as universally beneficial, we ask whether different explanation styles influence debugging performance and user perceptions for programmers with different assessed skill levels.

To support this evaluation, we first develop a short Python proficiency assessment focused on debugging-relevant competencies, including error comprehension, code understanding, fault localization, and error resolution. We then conduct a formative study to curate a set of buggy Python snippets and corresponding interpreter messages with appropriate difficulty for a crowdsourced debugging task. Finally, in a Prolific experiment with 103 participants, we compare standard Python interpreter messages against LLM-generated pragmatic and contingent rewrites. Participants attempt to repair a buggy Python snippet and then evaluate the message they received in terms of readability, cognitive load, and perceived tone.

Our study is guided by the following research questions:

\begin{framed}
\begin{itemize}
\item [\textbf{RQ1:}] How do skill-targeted LLM-generated Python error messages influence code-correction performance compared to standard interpreter messages across lower and higher assessed proficiency groups?

\item [\textbf{RQ2:}] How do programmers across lower and higher assessed proficiency groups perceive the readability, cognitive load, and tone of skill-targeted LLM-generated Python error messages compared to standard interpreter messages?

\end{itemize}
\end{framed}

Our results show a divergence between subjective experience and objective debugging performance. LLM-rewritten messages were rated more favorably than standard interpreter messages on several perceptual measures, with pragmatic messages receiving the strongest subjective evaluations. Participants perceived pragmatic messages as clearer, more helpful, and less cognitively demanding. However, these subjective benefits did not reliably translate into statistically significant improvements in objective debugging outcomes such as fix rate, time-to-fix, or number of attempts. Contingent messages also showed favorable subjective patterns, but their effects were less consistent than those of pragmatic messages.
This divergence reveals a human-AI complementarity gap in programming support: explanations that users experience as clearer and more helpful may not necessarily lead to better debugging performance. For HCI and human computation research, this finding highlights the need to evaluate AI-generated feedback not only by fluency or perceived helpfulness, but also by its effect on situated human problem solving. For programming-support systems, our results suggest that static skill-targeted rewriting may be insufficient on its own. Future systems may need to adapt explanations to the user's interaction state, repair trajectory, and specific error context, rather than relying only on prior skill classification.

This paper makes three contributions. First, we present a crowdsourced pipeline for designing and evaluating LLM-generated programming error explanations, including a debugging-oriented proficiency assessment and formative snippet-selection process. Second, we provide empirical evidence comparing standard Python interpreter messages with two styles of LLM-generated rewrites across objective debugging outcomes and subjective user perceptions. Third, we identify a gap between perceived explanation quality and debugging effectiveness, offering design implications for future human-AI programming support systems.

%% file: tex/2_Background.tex
\section{Background and Related Work}
\label{sec:background-related-work}

Programming error messages are a central feedback channel between software systems and programmers. They are produced automatically by compilers, interpreters, and runtime environments, but their usefulness depends on how users interpret and translate them into repair actions. This makes them a useful setting for studying human-AI complementarity: automated systems can produce diagnostic information, but human programmers must still understand, trust, and act on that information. We review work on programming error messages, LLM-based diagnostic explanations, skill-targeted feedback, and human evaluation of AI-generated programming support.

\subsection{Programming Error Messages as Human-Facing Diagnostics}
\label{subsec:compiler-interpreter-errors}

Programming error messages include both compile-time diagnostics, which are reported before a program runs, and runtime exceptions, which are reported during execution. Although these categories are often treated separately, the boundary is not strict across languages and execution environments. Many ecosystems combine ahead-of-time compilation, bytecode generation, just-in-time compilation, static analysis, and runtime diagnostics. We therefore use the broader term \textit{programming error messages} to refer to both compiler and interpreter feedback.

Prior work has repeatedly shown that programming error messages can be difficult for users to understand and act upon. Messages may be ambiguous, overly technical, poorly localized, or insufficiently explanatory, limiting their usefulness as repair guidance. Unclear or verbose messages can increase cognitive load, cause frustration, reduce productivity, and lead users toward ineffective repair attempts \cite{charlesEffectAutomatedError2023}. These problems are especially salient in educational contexts, where less experienced programmers may lack the conceptual vocabulary needed to map a diagnostic message to an underlying fault. However, more experienced programmers are also affected when messages are incomplete, misleading, or require substantial effort to interpret.

The form and difficulty of programming error messages also depend on language design and execution model. Functional languages\footnote{Functional programming (e.g., Haskell, F\#, Lisp) treats computation as the evaluation of mathematical functions, emphasizing immutability, declarative expressions, and higher-order functions \cite{hughes1989functional-matters, conception-evolution-fp}.} often produce complex type-related diagnostics, whereas object-oriented languages may surface errors related to inheritance, polymorphism, object state, or lifecycle management. Dynamically typed languages such as Python often produce runtime exceptions with limited local context. Because many modern languages are multi-paradigm\footnote{Python \cite{python-multi-paradigm}, for example, supports procedural, functional, and object-oriented constructs.}, no single message style is likely to serve all contexts. Effective feedback must therefore account for both the error context and the user's programming expertise.

\subsection{Improving Error Messages for Human Understanding}
\label{subsec:improving-error-messages}

Research on compiler and interpreter diagnostics has long argued that error messages should do more than report failure. They should help users identify the fault, understand its cause, and determine a plausible repair. \citet{compiler-error-messages-considered-unhelpful}'s survey of compiler error messages identifies recurring challenges, including excessive jargon, poor localization, insufficient contextual information, and lack of actionable guidance. Related work in computing education and programming languages has examined how learners interact with compiler messages, how diagnostics can be designed and evaluated, and why message clarity does not always translate into successful repair.

This literature motivates several design goals for improved programming error messages. Messages should be readable, using clear wording and avoiding unnecessary jargon; actionable, connecting the observed failure to a likely cause and useful next step; and appropriately scoped, providing enough information to support repair without disrupting the user's workflow. These goals align with broader HCI and learning-sciences work on feedback, scaffolding, and cognitive load. In programming contexts, the challenge is to support understanding without overwhelming users or interrupting the repair process.

\subsection{LLMs for Rephrasing and Explaining Diagnostics}
\label{subsec:llm-system-diagnostics}

LLMs create new opportunities for improving programming error messages. Unlike fixed diagnostic templates, they can combine the original traceback, surrounding code, and task-specific instructions to produce explanations that are more readable, contextual, or action-oriented. Recent work has explored LLMs for code explanation, repair, and diagnostic support, including rephrasing compiler or interpreter messages into more novice-friendly forms \cite{dcc-help, leinonen-codex, salmon2025debuggingerrormessagesllm, always-provide-context}. These studies show promise but also important limitations: generated explanations can be conceptually accurate yet inconsistent, novice-friendly yet incomplete, or clearer without reliably improving objective debugging outcomes \cite{dcc-help, leinonen-codex, not-the-silver-bullet}. Prompt structure and context inclusion also matter, with concise prompts, source-code context, and controlled examples improving clarity and fix specificity in some settings \cite{salmon2025debuggingerrormessagesllm, always-provide-context}. A large randomized study further suggests that GPT-generated messages can reduce repeated runs and additional attempts in authentic Python settings \cite{wang-rct}.

Together, this work suggests that LLM-generated diagnostics should not be treated as automatically beneficial. Model capability alone does not establish whether generated feedback improves human debugging: an explanation may be fluent, readable, and preferred by users while still failing to improve repair behavior. This makes LLM-generated diagnostics a human-centered evaluation problem rather than only a code-generation problem. Our work contributes to this line of research by evaluating both behavioral outcomes and subjective perceptions of LLM-rewritten Python error messages.

\subsection{Skill-Targeted Feedback and Explanation Style}
\label{subsec:skill-targeted-feedback}

Programmers with different levels of expertise may need different forms of feedback. Less experienced programmers may benefit from explanations that unpack the likely cause of an error, connect it to relevant programming concepts, and provide scaffolded guidance. More experienced programmers may prefer concise messages that preserve workflow and quickly identify the likely repair direction. This dis tinction is consistent with foundational work on scaffolding and cognitive load, where effective support depends on what the learner already knows and how much additional guidance the task requires \cite{wood1976role, sweller2011cognitive}. It is also consistent with the expertise reversal effect: instructional guidance that benefits less experienced learners can become redundant, inefficient, or even counterproductive for more knowledgeable users \cite{kalyuga2007expertise}.

In this paper, we examine two skill-targeted explanation styles. \textit{Pragmatic} messages are concise and task-oriented: they name the error, localize the likely cause, and provide a short hint toward repair. \textit{Contingent} messages provide more scaffolded guidance: they include additional context about the likely intent, possible misconception, and reasoning path. These styles are informed by work on actionable feedback, Bloom's taxonomy \cite{blooms-tax-revision, armstrong2010bloom}, and argumentation models such as claim-evidence-reasoning \cite{claim-evidence-reasoning} and Toulmin's model \cite{toulim-errors}. We use these frameworks to structure two explanation styles that may be differently suited to users with different assessed programming proficiency.

We do not evaluate a fully dynamic tutor that updates its feedback in response to user behavior over time. Instead, we evaluate \textit{skill-targeted rewriting}: the use of explanation styles designed for different proficiency needs. This provides an empirical basis for understanding whether static style differences are sufficient to improve debugging outcomes, or whether future systems need more interaction-aware adaptation.

Evaluating LLM-generated programming feedback requires more than testing whether a model can produce a plausible explanation. It requires human-centered evaluation of how users with different skill levels interpret, trust, and act on that feedback. Crowdsourcing offers a practical way to conduct such evaluations at scale, but it also requires careful calibration of both participants and tasks. In our study, crowdsourced input is used not only for final evaluation, but also to develop a debugging-oriented proficiency assessment and select code-repair tasks of appropriate difficulty.

%% file: tex/3_Experiment_Design.tex
\section{Study Design and Main Experiment}
\label{sec:study-design}

We used a multi-stage crowdsourced design to calibrate participant skill and debugging-task difficulty before evaluating LLM-generated programming error messages. This design reflects a core challenge in studying AI-generated feedback: explanation effects are difficult to interpret if participants' programming proficiency or task difficulty is poorly calibrated. Our pipeline consisted of three stages: (1) a debugging-oriented Python proficiency assessment, (2) a formative task-selection study, and (3) a main Prolific experiment comparing standard Python interpreter messages with two LLM-generated rewrite styles.

\begin{figure}[h!]
\centering
\includegraphics[width=\linewidth]{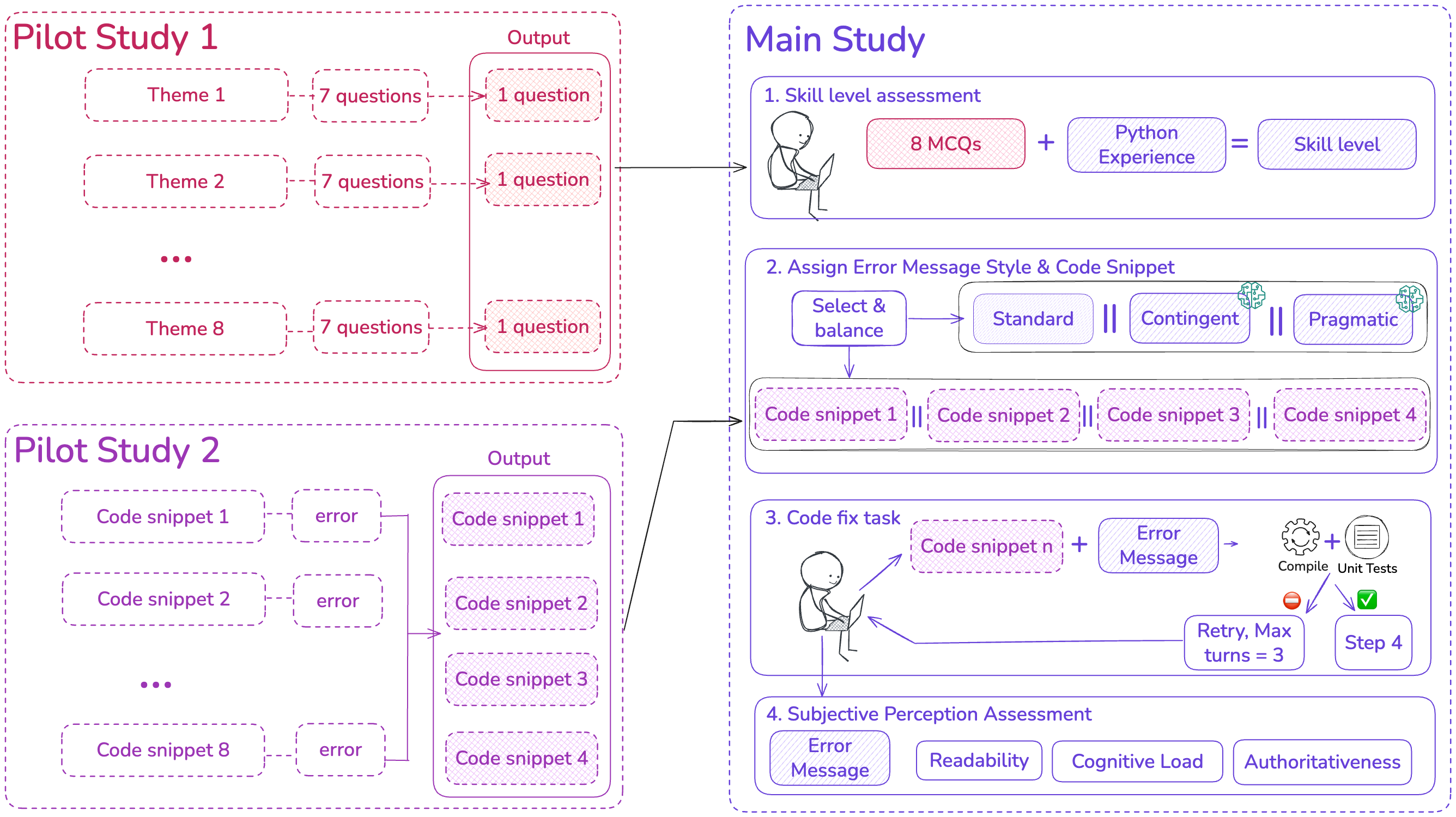}
\caption{Overview of the multi-stage study design. We first calibrated participant skill, then calibrated the debugging tasks, and finally evaluated standard and LLM-generated error messages in a controlled Prolific experiment.}
\label{fig:project-approach}
\end{figure}

\subsection{Crowdsourced Calibration}
\label{sec:crowdsourced-calibration}

\subsubsection{Participant Skill Calibration}
\label{sec:python-assessment}

Because our research questions compare explanation effects across programming skill levels, we developed a short Python proficiency assessment focused on debugging-relevant competencies. Self-reported experience alone may not reliably capture the skills required to interpret error messages and repair code. We therefore created a pool of 56 multiple-choice items spanning code comprehension, error-message interpretation, fault localization, and repair selection. In a pilot study, each participant answered 16 items sampled across these themes. We then selected eight items for the main experiment using item difficulty and point-biserial discrimination.

Participants in the main study were assigned to lower or higher assessed proficiency using both their score on the eight-item assessment and self-reported years of Python experience. We use the terms lower and higher assessed proficiency to avoid treating the categories as complete measures of programming expertise. Full details of the item pool, pilot procedure, pilot correlations, and final item statistics are provided in Appendix~\ref{appendix:skill-assessment}.

\begin{table}[h]
\centering
\small
\begin{tabularx}{\linewidth}{@{}c>{\centering\arraybackslash}X>{\centering\arraybackslash}X@{}}
\toprule
\textbf{MCQ score} & \textbf{Python experience} & \textbf{Skill group} \\
\midrule
$\geq 6$          & Any                       & Expert \\
$\leq 3$          & Any                       & Novice \\
$[4, 5]$          & $\geq 5$                  & Expert \\
$[4, 5]$          & $< 5$                     & Novice \\
\bottomrule
\end{tabularx}
\caption{Logic for assigning participants to skill level categories based on MCQ score and years of Python experience (YoE).}
\label{tab:skill-level-logic}
\end{table}
\subsubsection{Debugging Task Calibration}
\label{sec:task-calibration}

The main experiment required debugging tasks that were realistic enough to support meaningful repair behavior, but constrained enough for a crowdsourced study. We generated eight candidate Python snippets targeting common error families, including \texttt{NameError}, \texttt{TypeError}, and \texttt{SyntaxError}. Each snippet was self-contained, used only the Python standard library, and contained one primary fault. We manually checked that each candidate triggered the intended error and avoided errors that could lead to multiple simultaneous faults.

We then conducted a formative task-selection study in which participants rated each candidate snippet and standard interpreter message on code difficulty, fix difficulty, error-message mental demand, and error-message usefulness. Based on these ratings, we selected four snippets that provided variation in error type while remaining within a moderate difficulty range for the target participant population. The full selected snippets, standard interpreter messages, and task-selection survey items are provided in Appendix~\ref{appendix:debugging-tasks}.

\subsection{Message Conditions}
\label{sec:message-conditions}

The main experiment compared three error-message styles. The \textit{standard} condition showed the unmodified Python interpreter message. The \textit{pragmatic} condition used an LLM-generated rewrite designed to provide concise, action-oriented guidance. The \textit{contingent} condition used an LLM-generated rewrite designed to provide more scaffolded explanatory support, including additional context about the likely cause of the error and how the programmer might reason about it.

All participant-facing rewrites were generated using llama-3.1-8B-Instruct. The prompts included the code snippet, the standard interpreter message, and line-numbered context. Generation used zero-shot prompting with temperature set to 0 to improve consistency and reproducibility. Full prompt templates and generation details are provided in Appendix~\ref{appendix:prompts}. Figure~\ref{fig:error-message-examples} shows an example of the three message styles.

\begin{figure}[h!]
\centering
\begin{subfigure}[b]{0.31\linewidth}
    \centering
    \includegraphics[width=\linewidth]{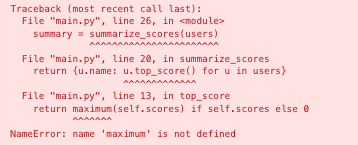}
    \caption{Standard interpreter message.}
\end{subfigure}
\hfill
\begin{subfigure}[b]{0.31\linewidth}
    \centering
    \includegraphics[width=\linewidth]{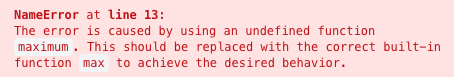}
    \caption{Pragmatic rewrite.}
\end{subfigure}
\hfill
\begin{subfigure}[b]{0.31\linewidth}
    \centering
    \includegraphics[width=\linewidth]{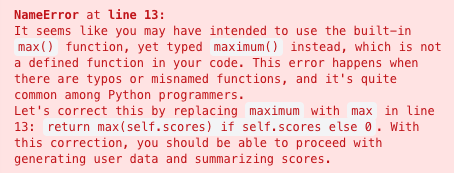}
    \caption{Contingent rewrite.}
\end{subfigure}
\caption{Example comparison of the standard Python interpreter message, pragmatic LLM rewrite, and contingent LLM rewrite.}
\label{fig:error-message-examples}
\end{figure}

\subsection{Main Prolific Experiment}
\label{sec:main-experiment}

The main study was deployed through a custom web application integrated with Prolific for recruitment and compensation. Participants were required to be at least 18 years old, fluent in English, have prior Python experience, have a Prolific approval rate above 90\%, and use a desktop or laptop computer. Participants from earlier pilot studies were excluded to avoid carryover effects.

After consent and instructions, participants completed the Python skill assessment and were assigned to one of the four calibrated debugging tasks and one of the three message styles. Assignment used constrained randomization to maintain balance across message styles and snippets. Each participant saw exactly one code snippet and one error-message style.

Participants attempted to repair the code in an embedded editor. A submission was counted as correct if it removed the targeted error and passed a hidden semantic test suite. Participants were allowed up to three attempts. After an incorrect attempt, the editor reset to the original code to avoid cumulative code drift. After the debugging task, participants rated the message they received in terms of readability, cognitive load, and perceived tone.

The final dataset included, N = 103 valid participants after quality control. Allocation was balanced across message style and assessed proficiency: 35 participants were assigned to the standard condition, 35 to the pragmatic condition, and 33 to the contingent condition. The sample included 38 participants in the lower assessed proficiency group and 65 in the higher assessed proficiency group. Detailed recruitment information, platform details, interface screenshots, quality-control procedures, survey items, and allocation tables are provided in Appendix~\ref{appendix:main-study-details}.

\subsection{Measures}
\label{sec:measures}

We measured both objective debugging outcomes and subjective message evaluations. Objective outcomes addressed RQ1 and included fix rate, number of attempts, Fix@k, and time-to-fix. Fix rate captured whether a participant successfully repaired the snippet within three attempts. Fix@k captured whether the participant first succeeded on attempt 1, 2, or 3. Time-to-fix measured elapsed time from task start until the first correct submission for participants who successfully repaired the snippet.

Subjective outcomes addressed RQ2 and included perceived readability, cognitive load, and tone. Readability was measured with items covering length, jargon, sentence structure, and vocabulary. Cognitive load was measured through intrinsic, extraneous, and germane load items. Tone was measured with a single item assessing whether the message felt respectful and reader-centered. Full item wording is provided in Appendix~\ref{appendix:main-study-details}.

\subsection{Participants}
\label{sec:participants}

We recruited participants through Prolific. Eligibility criteria required participants to be at least 18 years old, fluent in English, have prior Python experience according to Prolific screening, have a Prolific approval rate above 90\%, and use a desktop or laptop computer. Participants from earlier pilot studies were excluded to avoid carryover effects. The study was initially estimated to take 15--20 minutes; after a soft launch with 10 participants, the expected duration was updated to approximately 24 minutes. Final completion times averaged 26 minutes, with a median of 23 minutes. Participants were compensated at £9/hour.

A total of 103 participants met all quality-control criteria and were included in the final analysis. Quality control relied on logged browser focus events, copy/paste blocking with browser alerts, task completion times, and submission patterns. Triggered cases were manually reviewed, and participants showing random typing, repeated unchanged submissions, or clear low-effort behavior were excluded.

The final sample included 38 participants in the lower assessed proficiency group and 65 in the higher assessed proficiency group. Participants were distributed across message-style conditions as follows: 35 received the Standard interpreter message, 35 received the Pragmatic LLM rewrite, and 33 received the Contingent LLM rewrite. We report available age, gender, and nationality metadata descriptively in Appendix~\ref{appendix:main-study-details}, but do not conduct demographic subgroup analyses because they are outside the scope of our research questions.

%% file: tex/5_Results.tex


\section{Results}
\label{sec:results}

We report results from the main Prolific experiment with 103 valid participants. Each participant attempted one calibrated Python debugging task and received one of three message styles: \textit{Standard}, \textit{Pragmatic}, or \textit{Contingent}. We first report objective debugging outcomes addressing RQ1, followed by subjective message evaluations addressing RQ2.

\subsection{Objective and Subjective Outcomes}
\label{sec:prolific-main-study-results}

\noindent

The participants of our main study (\textit{N}=\,103) attempted to fix one of four buggy Python snippets (balanced across conditions) and were randomly assigned to one of three message styles: \textit{Standard} (baseline), \textit{Pragmatic}, or \textit{Contingent}. We first report \emph{objective performance}, such as fix rates, Fix@\{1,2,3\}, and time-to-fix in Section \ref{subsec:fixrate-objective}. We then report \emph{subjective perceptions}, namely readability, perceived cognitive load, and perceived authoritativeness in Section \ref{subsec:subjective-perception}. Finally, we also perform \emph{exploratory analyses} that investigate additional potentially unexplored patterns in Section \ref{subsec:exploratory}, in addition to the main results. 

As a reminder for interpreting the results in the following sections, participants were classified according to our assignment logic, which combined their performance on the skill-assessment survey with their self-reported Python YoE. Out of the 103 participants, 65 were categorized as experts and 38 as novices. The distribution of participants across skill levels and snippets is provided in Table~\ref{tab:allocation}.

\subsubsection{Fix Rate, Fix@k, and Time-to-Fix}
\label{subsec:fixrate-objective}

For measuring participants' objective performance on the code fix task, across different subgroups we record the following metrics:
\begin{itemize}
  \item \textbf{Fix rate}: proportion of participants whose final submission is fully correct.
  \item \textbf{Fix@1/2/3}: correctness achieved within the first, second, or third submission attempt.
  \item \textbf{Time-to-fix}: total seconds from code fix task start until the first correct submission.
\end{itemize}

\paragraph{\textbf{Fix Rate:}}

Table \ref{tab:overall-fix-by-snippet} reports overall fix rates by snippet across all participants, and Table \ref{tab:fix-by-style-skill} breaks them down by message style and skill level. Fix rate, and therefore difficulty, varies significantly across snippets, with snippet A showing the lowest success rate. This outcome is expected because the standard interpreter output for snippet A fails to reveal the true root cause, and actually points to the non-offending line\footnote{Interested readers can refer to Listing \ref{lst:fantastic-four-snippet-a} and Figure \ref{fig:snippet-a-error-message} in Appendix \ref{appendix:fantastic four code snippets}, which present the code and the corresponding standard error message for the least solved buggy code snippet in our study (when being shown the standard interpreter output).}. 

Looking more in-depth, Tables \ref{tab:fix-rates-set-novice} and \ref{tab:fix-rates-set-experts} report fix rates by message style and snippet ID for novices and experts. For snippet A, LLM-enhanced PEMs, which do correctly identify the offending line and root cause, generally increase fix rates, with the pragmatic style among novices as the main exception. However, across snippets, style effects vary widely, so we refrain from making any conclusions.

\begin{table}[h]
  \centering
  \begin{tabular}{lcccc}
    \toprule
    & \textbf{A} & \textbf{B} & \textbf{C} & \textbf{D} \\
    \midrule
    Fix Rate & 53\% ($n=26$) & 84\% ($n=26$) & 64\% ($n=25$) & 69\% ($n=26$)\\
    \bottomrule
  \end{tabular}
  \caption{Fix rate (\%) by snippet ID, across all participants; $n$ = the total number of participants assigned the given snippet ID.}
  \label{tab:overall-fix-by-snippet}
\end{table}

\begin{table}[h]
  \centering
  \begin{tabular}{lccc}
    \toprule
                 & \textbf{Standard} & \textbf{Pragmatic} & \textbf{Contingent} \\
    \midrule
    Novice ($n=38$) & 46\% & 54\% & 42\% \\
    Expert ($n=65$) & 68\% & 82\% & 90\% \\
    \bottomrule
  \end{tabular}
    \caption{Fix rate (\%) by message style and skill group (all snippets pooled).}
  \label{tab:fix-by-style-skill}
\end{table}

\begin{table}[h]
\centering
\begin{tabular}{lcccc}
\toprule
 & \textbf{A} & \textbf{B} & \textbf{C} & \textbf{D} \\
\midrule
\textbf{Standard}   & 33\% ($n=3$) & 67\% ($n=3$) & 33\% ($n=3$) & 50\% ($n=4$) \\
\textbf{Pragmatic}  & 0\%  ($n=3$) & 100\% ($n=4$) & 67\% ($n=3$) & 33\% ($n=3$) \\
\textbf{Contingent} & 67\% ($n=3$) & 67\% ($n=3$) & 0\% ($n=3$) & 33\% ($n=3$) \\
\bottomrule
\end{tabular}
\caption{Fix rate (\%) by message style and snippet ID for novice participants. The number $n$, in parentheses, represents the total number of participants assigned the given combination of snippet ID and error message style.}
\label{tab:fix-rates-set-novice}
\end{table}

\begin{table}[h]
\centering
\begin{tabular}{lcccc}
\toprule
 & \textbf{A} & \textbf{B} & \textbf{C} & \textbf{D} \\
\midrule
\textbf{Standard}   & 33\% ($n=6$) & 80\% ($n=5$) & 67\% ($n=6$) & 100\% ($n=5$) \\
\textbf{Pragmatic}  & 83\% ($n=6$) & 100\% ($n=5$) & 80\% ($n=5$) & 67\% ($n=6$) \\
\textbf{Contingent} & 80\% ($n=5$) & 83\% ($n=6$) & 100\% ($n=5$) & 100\% ($n=5$) \\
\bottomrule
\end{tabular}
\caption{Fix rate (\%) by message style and snippet ID for expert participants. The number $n$, in parentheses, represents the total number of participants assigned the given combination of snippet ID and error message style.}
\label{tab:fix-rates-set-experts}
\end{table}

\paragraph{\textbf{Fix@k:}}

\begin{figure}[h!]
    \centering
    \includegraphics[width=0.70\linewidth]{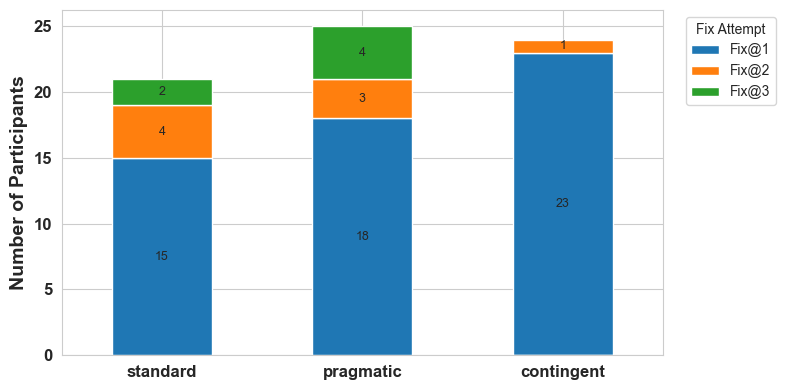}
    \caption{Fix@k, for $k\in{1, 2, 3}$, when varying error message style across all snippets and skill levels.}
    \label{fig:fix-at-k}
\end{figure}

We initially introduced the Fix@k metric to quantify how many attempts participants needed to repair each buggy snippet. Our aim was to test whether error message style influenced repeated submissions across the full sample and within novice and expert subgroups. In practice, results were situated at the extremes. Most participants either fixed the error on the first try, or did not succeed at all within the three allowed attempts. Because Fix@k showed limited variation (Figure~\ref{fig:fix-at-k}), we could not conduct meaningful further analysis, so Fix@k remains largely unexplored in this study. A visual reading of the figure nevertheless suggests that pragmatic and contingent messages may yield higher fix rates, in general, than the baseline, since they outperform standard at $k=1$. For contingent messages in particular, outcomes often cluster at either ``fixed on the first attempt'' or ``not at all within three attempts''. Therefore, interpreting low Fix@2 and Fix@3 counts requires care, since they can result from either strong first-attempt performance or a large tail of unresolved cases. Future studies should relax the bound on $k$ (within reason) to better visualize the average number of attempts per error-message style until successful code repair.

\paragraph{\textbf{Time-to-Fix:}}

\begin{figure}[h!]
    \centering
    \includegraphics[width=0.5\linewidth]{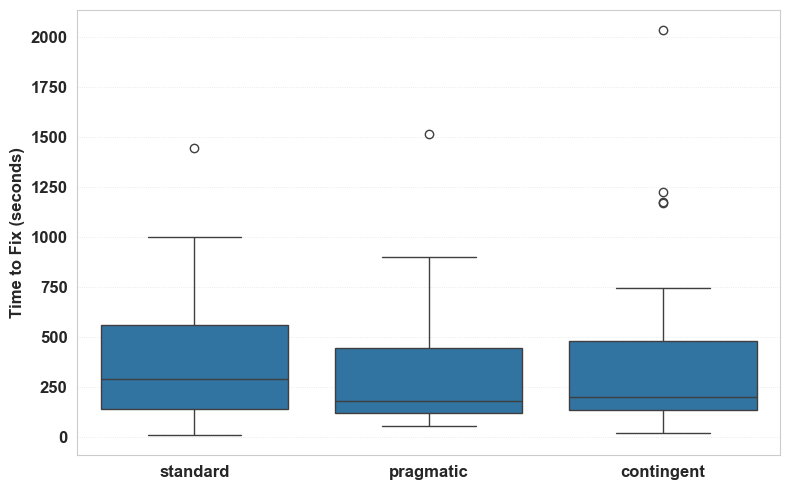}
    \caption{Box plot for time-to-fix (seconds) by message style. Time-to-fix sums all attempts until the first success.\\Standard ($M=415.69, SD=377.75$); Pragmatic ($M=324.56, SD=333.12$); Contingent ($M=423.85, SD=501.47$).}
    \label{fig:time-to-fix-all-skill-levels}
\end{figure}

\begin{figure}[h!]
  \centering
  \begin{subfigure}[b]{0.48\linewidth}
    \centering
    \includegraphics[width=\linewidth]{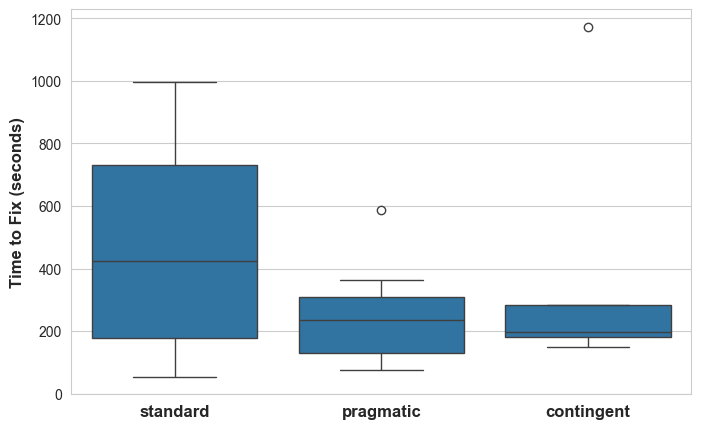}
    \caption{Time-to-fix box plot for novices, by message style.\\Standard ($M=471.58, SD=375.0$); Pragmatic ($M=253.71, SD=178.52$); Contingent ($M=396.41, SD=437.06$).}
    \label{fig:prolific-time-to-fix-novices}
  \end{subfigure}
  \hfill
  \begin{subfigure}[b]{0.48\linewidth}
    \centering
    \includegraphics[width=\linewidth]{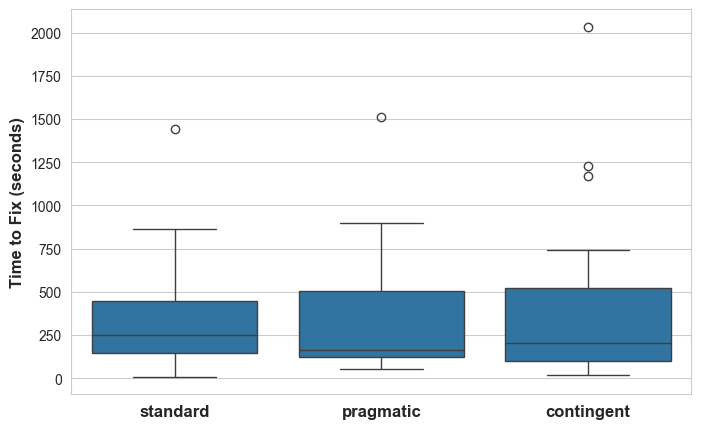}
    \caption{Time-to-fix box plot for experts, by message style.\\Standard ($M=393.34, SD=389.57$); Pragmatic ($M=352.13, SD=377.56$); Contingent ($M=431.08, SD=527.85$).}
    \label{fig:prolific-time-to-fix-experts}
  \end{subfigure}
  \caption{Box plots for time-to-fix (seconds) by message style and skill level; left plot for novices, and right plot for experts.}
  \label{fig:prolific-time-to-fix-skill-level-grouping}
\end{figure}

As for the time-to-fix metric, Figure~\ref{fig:time-to-fix-all-skill-levels} shows that, averaged across skill levels, participants fixed errors fastest when shown pragmatic messages. Contingent messages also reduced time-to-fix relative to the baseline, but less than pragmatic, which is unsurprising given their greater verbosity and the extra reading time required. However, when split by skill level (see Figure~\ref{fig:prolific-time-to-fix-skill-level-grouping}), both LLM-enhanced PEMs tend to speed fixes versus standard. With few novice successes (standard $n{=}6$, pragmatic $n{=}7$, contingent $n{=}5$), statistical tests found no significant effects, though the current pattern seems to favor LLM explanations. As for experts, results remain inconclusive.

\subsubsection{Subjective Evaluations}
\label{subsec:subjective-perception}

\begin{figure}[h!]
  \centering
  \begin{subfigure}[b]{0.4\linewidth}
    \centering
    \includegraphics[width=\linewidth]{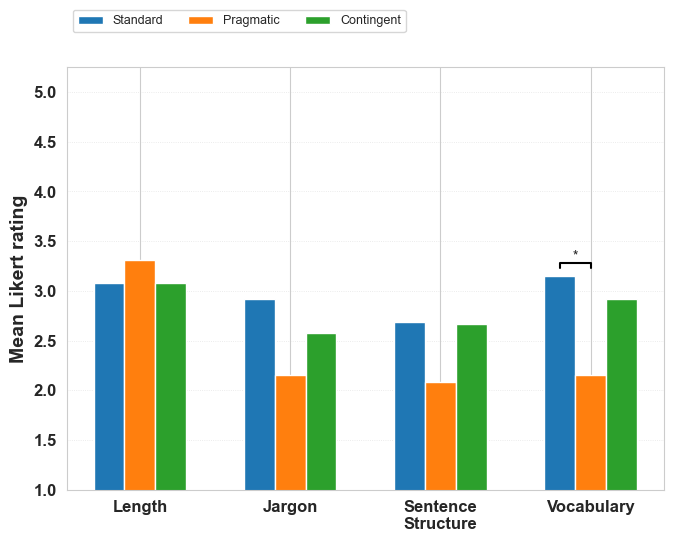}
    \caption{Readability metrics (Novices).}
    \label{fig:read-nov}
  \end{subfigure}
  \hfill
  \begin{subfigure}[b]{0.4\linewidth}
    \centering
    \includegraphics[width=\linewidth]{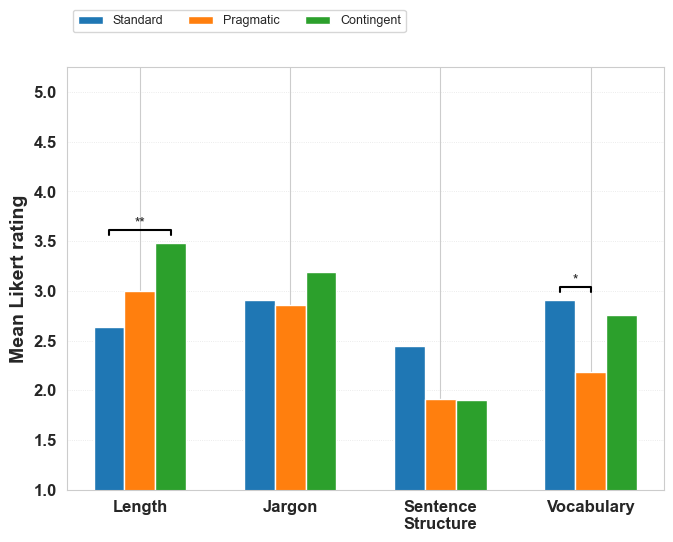}
    \caption{Readability metrics (Experts).}
    \label{fig:read-exp}
  \end{subfigure}
    \caption{Mean ratings for readability by message style and skill group. Brackets with asterisks indicate pairwise differences from Dunn’s post hoc tests (Holm-adjusted) following Kruskal-Wallis: * ($p<0.1$), ** ($p<0.05$), and *** ($p<0.01$).}
  \label{fig:subjective-readability-split-by-skill-level}
\end{figure}

\begin{figure}[h!]
  \centering
  \begin{subfigure}[b]{0.4\linewidth}
    \centering
    \includegraphics[width=\linewidth]{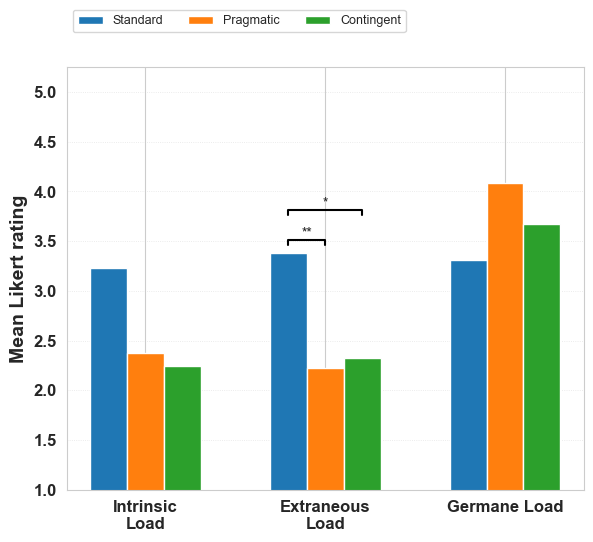}
    \caption{Cognitive load metrics (Novices).}
    \label{fig:load-nov}
  \end{subfigure}
  \hfill
  \begin{subfigure}[b]{0.4\linewidth}
    \centering
    \includegraphics[width=\linewidth]{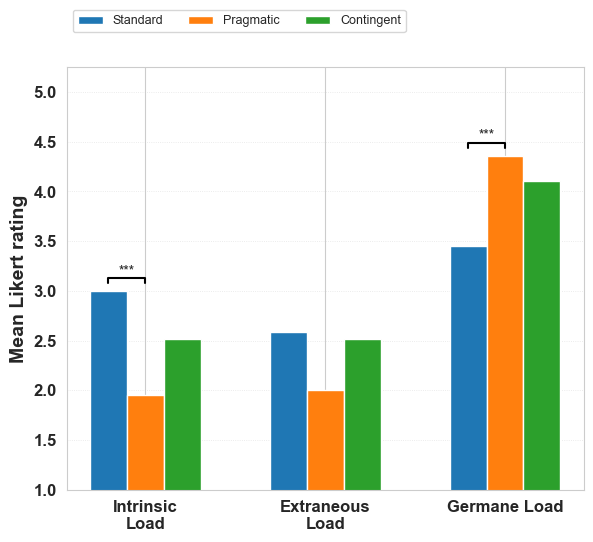}
    \caption{Cognitive load metrics (Experts).}
    \label{fig:load-exp}
  \end{subfigure}
    \caption{Mean ratings for cognitive load by message style and skill group. Brackets with asterisks indicate pairwise differences from Dunn’s post hoc tests (Holm-adjusted) following Kruskal-Wallis: * ($p<0.1$), ** ($p<0.05$), and *** ($p<0.01$).}
  \label{fig:subjective-cognitive-load-split-by-skill-level}
\end{figure}

\begin{figure}[h!]
  \centering
  \begin{subfigure}[b]{0.4\linewidth}
    \centering
    \includegraphics[width=\linewidth]{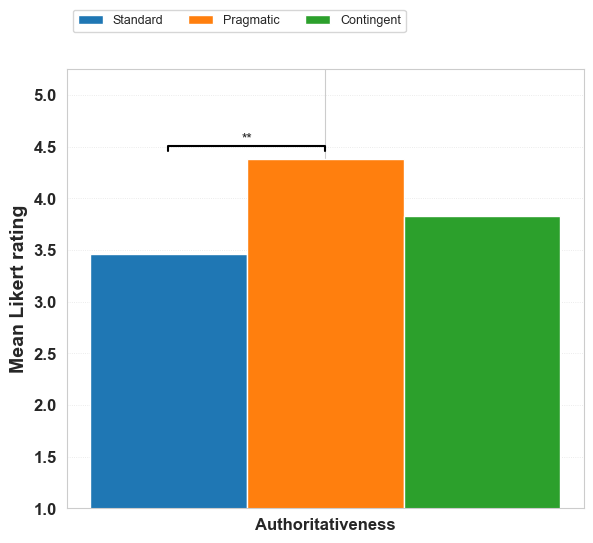}
    \caption{Authoritativeness metric (Novices).}
    \label{fig:auth-nov}
  \end{subfigure}
  \hfill
  \begin{subfigure}[b]{0.4\linewidth}
    \centering
    \includegraphics[width=\linewidth]{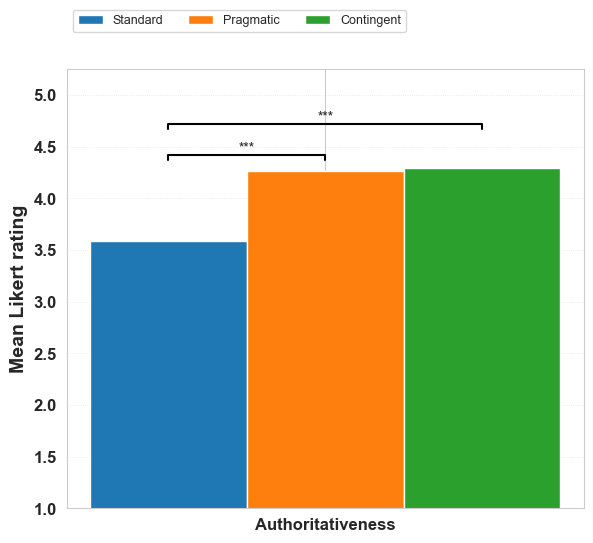}
    \caption{Authoritativeness metric (Experts).}
    \label{fig:auth-exp}
  \end{subfigure}
    \caption{Mean ratings for authoritativeness by message style and skill group. Brackets with asterisks indicate pairwise differences from Dunn’s post hoc tests (Holm-adjusted) following Kruskal-Wallis: * ($p<0.1$), ** ($p<0.05$), and *** ($p<0.01$).}
  \label{fig:subjective-authoritativeness-split-by-skill-level}
\end{figure}

Figures \ref{fig:subjective-readability-split-by-skill-level}, \ref{fig:subjective-cognitive-load-split-by-skill-level}, \ref{fig:subjective-authoritativeness-split-by-skill-level}, show mean ratings by error message style and skill level. For novices, Holm-adjusted Dunn tests indicate lower \emph{extraneous load} for Pragmatic vs.\ Standard (\(p = 0.039\)) and also for Contingent vs.\ Standard (\(p = 0.054\)). The results also report lower \emph{authoritativeness} for Pragmatic vs.\ Standard (\(p = 0.011\)), while \emph{vocabulary} also shows a trend favoring Pragmatic vs.\ Standard (\(p = 0.087\)), meaning Pragmatic messages are generally seen as using simpler vocabulary. 

For experts, Pragmatic vs.\ Standard is associated with lower \emph{intrinsic} (\(p = 0.0049\)) and higher \emph{germane} load (\(p = 0.0043\)), and both Pragmatic and Contingent are rated less \emph{authoritative} than the Standard ones (\(p = 0.0097\) in both contrasts). Experts also perceive messages as more verbose for Contingent vs.\ Standard (\(p = 0.035\)), with a trend toward simpler \emph{vocabulary} for Pragmatic vs.\ Standard (\(p = 0.088\)). Overall, adaptive style, especially Pragmatic, are consistently rated by experts as lighter in cognitive load, less authoritative, and more useful than the Standard baseline. Contingent error messages are also, on average, rated better by experts across almost all metrics, yet their increased verbosity seems to impact their overall usefulness.

\begin{figure}[h!]
    \centering
    \includegraphics[width=0.8\linewidth]{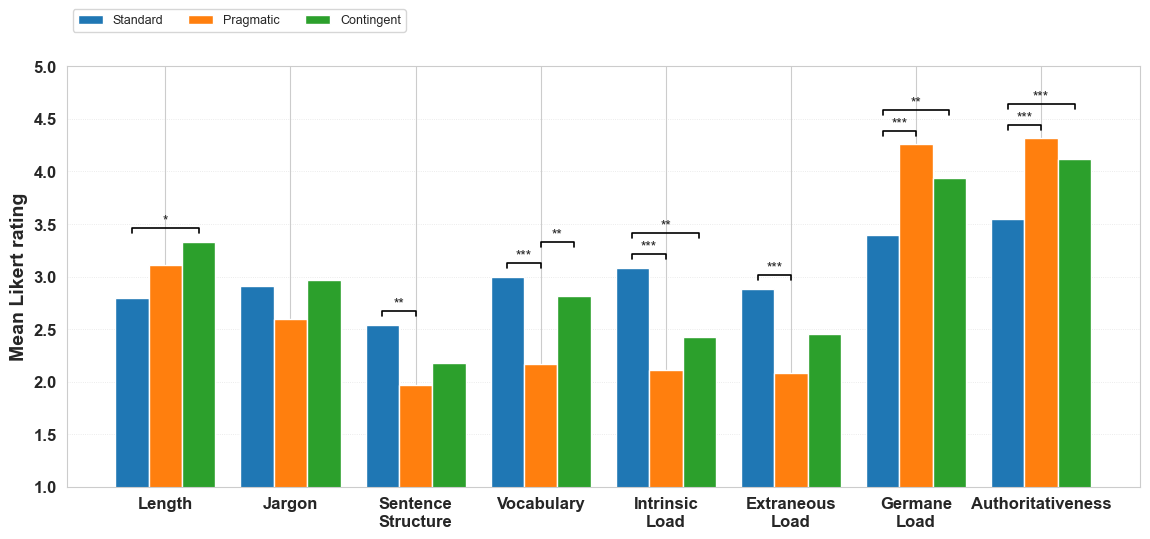}
    \caption{Mean ratings for readability, cognitive load, and authoritativeness by message style. Brackets with asterisks indicate pairwise differences from Dunn’s post hoc tests (Holm-adjusted) following Kruskal-Wallis: * ($p<0.1$), ** ($p<0.05$), and *** ($p<0.01$).}
    \label{fig:subjective-ratings-all}
\end{figure}

Figure~\ref{fig:subjective-ratings-all} reports mean ratings for the full sample, aggregated only by message style. Similarly, we also perform Kruskal-Wallis tests followed by Holm-adjusted Dunn comparisons where necessary, and the results indicate several pairwise differences. 

In terms of \textit{readability}, differences appear for \emph{sentence structure} between Pragmatic and Standard (\(p=0.0218\)), for \emph{vocabulary} between Pragmatic and Standard (\(p=0.0063\)) and Pragmatic vs.\ Contingent (\(p=0.0423\)), and a trend for \emph{length} between Standard and Contingent (\(p=0.0949\)). 

For \textit{cognitive load}, we observe differences in \emph{intrinsic load} for Pragmatic vs.\ Standard (\(p=0.0017\)) and Contingent vs.\ Standard (\(p=0.0431\)). In \emph{extraneous load} we notice differences only for Pragmatic vs.\ Standard (\(p=0.0064\)). We also reported differences in \emph{germane load}, both for Pragmatic vs.\ Standard (\(p=0.0013\)) and Contingent vs.\ Standard (\(p=0.0488\)). 

Finally, in terms of \textit{authoritativeness}, both Pragmatic and Contingent differ from the Standard error message style (\(p=0.0001\) and \(p=0.0032\), respectively). Overall, the most frequent contrasts involve Pragmatic (and, in several cases, Contingent) relative to the Standard, consistent with adaptive styles differing from the baseline interpreter output on readability, perceived load, and tone.

For the previously mentioned statistical analysis, we use the Kruskal-Wallis test with Dunn's pairwise comparisons (Holm-adjusted) because our outcomes are Likert ratings (i.e., ordinal, bounded), and after performing validation checks we saw that the data was not normally distributed. Kruskal-Wallis is an alternative to ANOVA that handles such data well and tolerates unequal group sizes. When the omnibus test indicates a difference, Dunn's test tells us which styles differ. Holm's adjustment controls false positives across the three pairwise checks per metric while being less conservative (and more powerful) than an alternative like Bonferroni.

\subsubsection{Exploratory Analysis}
\label{subsec:exploratory}

\begin{figure}[h!]
  \centering
  \begin{subfigure}[b]{0.48\linewidth}
    \centering
    \includegraphics[width=\linewidth]{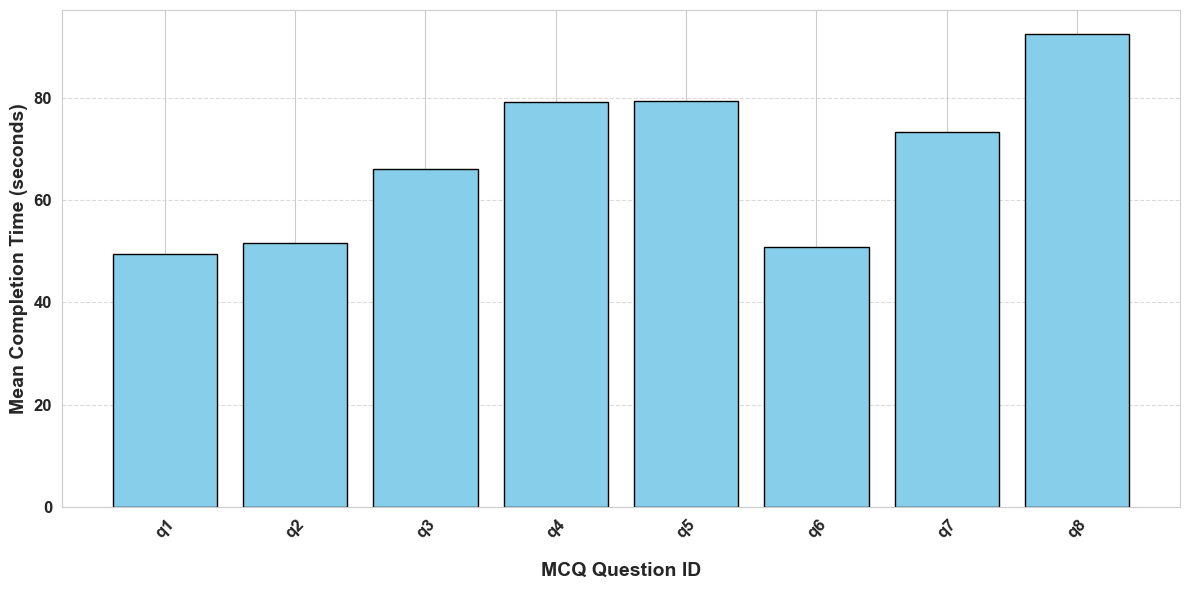}
    \caption{Mean completion/answering time in seconds for each MCQ in the Prolific study.}
    \label{fig:prolific-mcq-mean-completion-time}
  \end{subfigure}
  \hfill
  \begin{subfigure}[b]{0.48\linewidth}
    \centering
    \includegraphics[width=\linewidth]{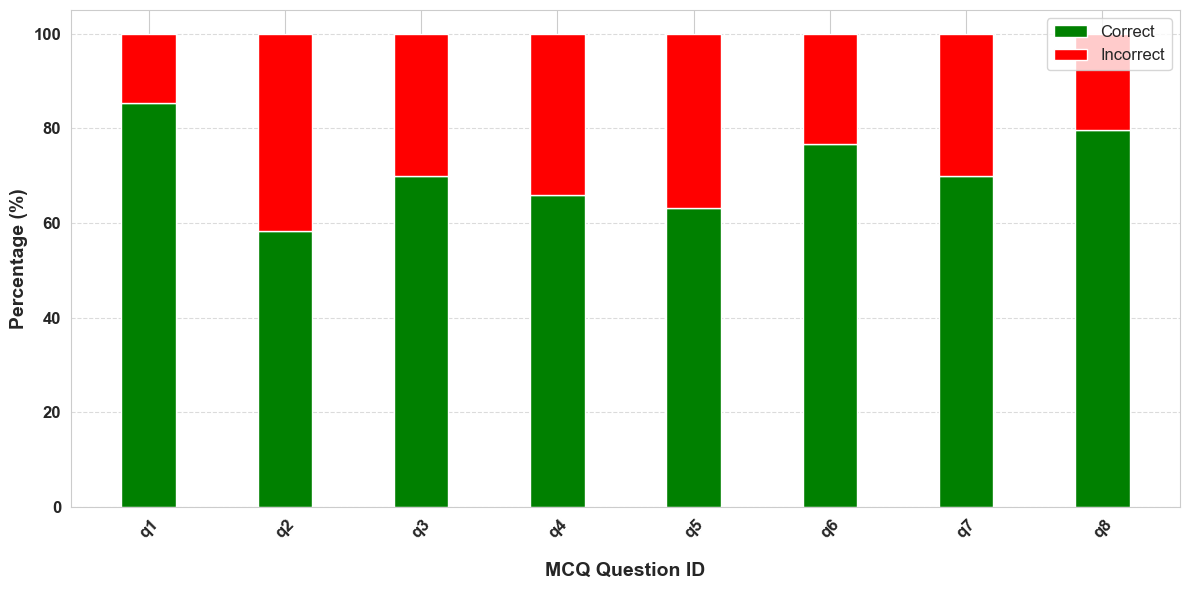}
    \caption{Percentage of correct and incorrect answers for each MCQ in the Prolific study.}
    \label{fig:prolific-correct-incorrect-mcq}
  \end{subfigure}
  \caption{Statistics, such as mean completion time and correctness percentage pertaining to the MCQs in the Prolific study.}
  \label{fig:prolific-mcq-statistics}
\end{figure}

In addition to the previous results, we also performed additional exploratory analysis to see whether we could find uncovered patterns. For instance, Figure \ref{fig:prolific-mcq-statistics} depicts statistics about the multiple-choice questions that compose the Python skill assessment in the first part of the Prolific study, where we are reporting mean completion time and question correctness percentage. On average, we noticed that correct answers tend to require more time to select, indicating that questions were mostly answered thoughtfully and legitimately, rather than at random. 

Another aspect we wanted to verify was whether the piloted Python skill assessment survey reliably reflects true proficiency and predicts debugging performance in the main Prolific study. To that end, we tested whether performance on the survey, measured as the number of MCQs answered correctly, relates to fixing the assigned code snippet. A point-biserial correlation showed a moderate positive association, $r_{pb}=0.368$ with $p=0.0001$, indicating that higher MCQ scores correspond to a greater likelihood of producing a correct fix.

\begin{figure}[h!]
  \centering
  \includegraphics[width=0.6\linewidth]{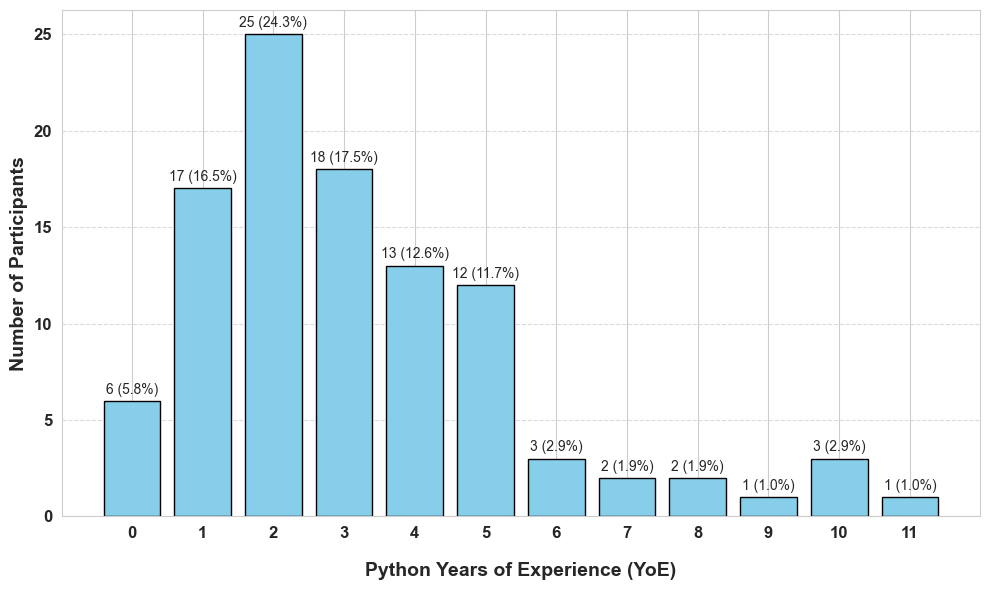}
  \caption{Prolific participants' self-reported Python YoE distribution.}
  \label{fig:prolific-yoe-distribution}
\end{figure}

%% file: tex/6_discussion.tex
\section{Discussion}
\label{sec:discussion}

Our study examined whether LLM-rewritten Python error messages can support code repair and improve users' perceptions of programming diagnostics. Overall, the results show a consistent divergence between subjective and objective outcomes. Participants rated LLM-generated messages, especially Pragmatic messages, more favorably than standard Python interpreter messages on readability, cognitive load, and tone. However, these subjective advantages did not reliably translate into improved debugging performance. This section discusses the implications of that divergence, the role of skill-targeted explanation styles, and the limitations of static rewriting as a form of programming support.

\subsection{Perceived Explanation Quality Does Not Guarantee Better Debugging}
\label{subsec:discussion-subjective-objective-gap}

The clearest finding from our study is that participants experienced LLM-rewritten messages as better than standard interpreter messages, but this did not produce reliable objective performance gains. Pragmatic messages were perceived as clearer, less cognitively demanding, and less authoritative than standard messages. Contingent messages also improved several subjective ratings, although less consistently and with signs of increased verbosity. In contrast, objective outcomes such as fix rate, time-to-fix, and Fix@k were more variable and did not provide strong evidence that rewritten messages reliably improved repair performance.

This gap is important for the design and evaluation of AI-generated programming support. A message that feels clearer may still fail to help the user locate the fault, identify the correct repair, or implement that repair successfully. Debugging requires more than interpreting the diagnostic message: users must connect the message to the program state, reason about the intended behavior, and modify the code correctly. LLM-generated explanations may reduce the friction of reading an error message without resolving these downstream reasoning and repair challenges.

For human-AI systems, this suggests that perceived helpfulness should not be treated as a proxy for task effectiveness. Subjective evaluations remain important because they capture whether users find the system understandable, respectful, and worth engaging with. However, they need to be interpreted alongside behavioral measures. Our results, therefore, support a more cautious evaluation standard for LLM-generated explanations: fluency, readability, and user preference are necessary but insufficient indicators of effective support.

\subsection{Skill-Targeted Rewriting Was Useful, But Not Sufficiently Adaptive}
\label{subsec:discussion-skill-targeted}

Our results also complicate the assumption that more scaffolded explanations are necessarily better for less experienced programmers. Contingent messages were designed to provide additional context and reasoning support, but they did not consistently benefit the lower assessed proficiency group. One explanation is that scaffolding can introduce additional reading and interpretation costs. For users still struggling to understand the code, a longer explanation may increase cognitive effort even if it contains useful information. This interpretation is consistent with the broader concern that instructional support must be matched not only to a user's general skill level, but also to their immediate task state.

Pragmatic messages produced the most consistent subjective benefits. Their concise and action-oriented form may have helped participants extract a likely repair direction without requiring them to process a longer explanation. This suggests that, for short debugging tasks, reducing friction may sometimes be more valuable than adding pedagogical detail. However, pragmatic messages also did not produce reliable objective gains, which indicates that concise explanations alone are not enough to ensure successful repair.

These findings clarify the limits of our current approach. We evaluate skill-targeted rewriting: explanation styles designed for different assessed proficiency groups. We do not evaluate a fully dynamic tutor that adapts to the user's behavior during debugging. 
A participant's assessed Python proficiency may not reflect their local understanding of a specific bug, their familiarity with the underlying concept, or their moment-to-moment repair strategy. Future systems may therefore need to adapt explanations based on interaction signals such as repeated failed attempts, time spent before editing, edit location, error recurrence, or requests for additional detail. Static skill categories are a useful starting point, but they are likely too coarse to support robust personalization.

 \subsection{Implications for HCOMP and AI Feedback Evaluation}
 \label{subsec:discussion-hcomp}

 Our study also illustrates the value of crowdsourced calibration for evaluating AI-generated programming feedback. Our evaluation did not begin with the final Prolific experiment alone. Human input was used to pilot a debugging-oriented proficiency assessment, calibrate candidate code snippets, and evaluate message perceptions after repair attempts. This multi-stage process helped reduce two common threats in crowdsourced programming studies: poorly measured participant skill and poorly calibrated task difficulty.
 At the same time, the results show that calibration cannot fully remove heterogeneity. Snippet difficulty varied substantially, and some message-style effects appeared sensitive to the specific error context. This suggests that future evaluations of AI-generated programming support should avoid treating “the error message” as a single uniform intervention. The same explanation style may help in one error context and not another. Evaluations should therefore report effects by error type, task difficulty, and user proficiency where possible, while being cautious about overinterpreting small subgroups.


\subsection{Limitations and Future Work}
\label{subsec:limitations-future-work}

Our study has several limitations. First, our proficiency assessment was designed and piloted for this study, but it was not validated at large scale against external measures of Python expertise. The lower and higher assessed proficiency groups should therefore be interpreted as task-aligned operational categories rather than complete measures of programming expertise. Future work should validate such assessments against richer indicators, including course performance, instructor ratings, prior programming tasks, or longitudinal debugging behavior.

Second, our debugging tasks were constrained by the requirements of a crowdsourced study. The four Python snippets were short, synthetic, and designed to contain a single primary fault. This made the study feasible and controlled, but it limits ecological validity. Real-world debugging often involves larger files, multiple interacting faults, project-specific dependencies, build systems, and incomplete contextual knowledge. Effects observed in short web-based tasks may therefore differ from effects in authentic IDE or classroom settings.

Third, our system used static message styles generated from fixed prompt templates. We used \texttt{llama-3.1-8B-Instruct} with temperature 0 to support reproducibility and stable generation, but did not systematically vary model size, decoding settings, prompt structure, or error-specific templates. Different models or prompts may produce different tradeoffs between correctness, verbosity, and usefulness. 

Future work should test skill-targeted programming explanations in more ecologically valid settings. One direction is classroom deployment, where true novice programmers use rewritten error messages during programming assignments, and learning outcomes can be measured over time. Another direction is IDE-based evaluation, where explanations can be shown alongside standard diagnostics during authentic debugging workflows. Such studies could measure not only the fix rate and time-to-fix, but also edit trajectories, repeated error recurrence, hint requests, and delayed learning outcomes. Future systems should move beyond static skill-targeted rewriting toward interaction-aware support. Rather than assigning users to a fixed explanation style, systems could progressively disclose more detail when users struggle, offer concise summaries when users move quickly, or adapt explanations based on repeated failures and repair behavior. Such systems should also be evaluated for risks, including over-reliance, false confidence, and the possibility that fluent but incorrect explanations reinforce misconceptions. Our findings suggest that LLM-generated error messages are promising, but their value depends on careful human-centered evaluation and adaptive design.

%% file: tex/7_conclusion.tex
\section{Conclusion}
\label{sec:conclusion}

This paper examined whether LLM-generated rewrites of Python interpreter error messages can improve debugging support for programmers with different assessed proficiency levels. We developed a multi-stage crowdsourced pipeline that combined a debugging-oriented Python proficiency assessment, calibrated buggy code snippets, and a controlled Prolific experiment comparing standard interpreter messages with two LLM-generated styles: \textit{Pragmatic}, a concise and action-oriented rewrite, and \textit{Contingent}, a more scaffolded explanation intended to provide additional reasoning support.
Our findings show a clear divergence between subjective and objective outcomes. Participants rated LLM-rewritten messages, especially Pragmatic messages, as more readable, less cognitively demanding, and less authoritative than standard interpreter output. However, these perceptual benefits did not reliably translate into statistically significant improvements in objective debugging outcomes such as fix rate, time-to-fix, or attempts. Contingent messages provided additional explanatory context, but their greater verbosity may have reduced their usefulness in short repair tasks, particularly for lower assessed proficiency participants.

These results suggest that LLM-generated programming explanations should not be evaluated only by fluency, readability, or user preference. Explanations that feel clearer may still fail to improve repair behavior. For human-AI programming support, this points to the need for evaluation methods that combine subjective perception, behavioral outcomes, task calibration, and user proficiency assessment. More broadly, our study suggests that static skill-targeted rewriting is a useful starting point but may be insufficient for robust debugging support. Future systems may need to adapt explanations dynamically to the user's interaction state, repair trajectory, and error context.

%% file: appendix/appendix_study_design.tex
\newpage
\appendix

\section{Python Proficiency Assessment}
\label{appendix:skill-assessment}

This appendix provides additional details on the Python proficiency assessment used to assign participants to lower and higher assessed proficiency groups in the main study. The assessment was designed to capture debugging-relevant competencies rather than general programming identity or self-confidence.

\subsection{Assessment Themes}

We created an initial pool of 56 multiple-choice items, with seven items across eight themes. These themes were selected to cover competencies involved in interpreting error messages, understanding code, locating faults, and selecting repairs.

\begin{itemize}
\item \textbf{General Programming Error Understanding}: Items assessing general knowledge about programming errors independent of any specific language.
\item \textbf{Python-Specific Error Understanding}: Items focused on Python syntax, semantics, indentation, keywords, and language-specific constructs.
\item \textbf{Code Reading and Understanding}: Items assessing participants' ability to interpret Python code and reason about its behavior.
\item \textbf{Error Message Comprehension}: Items assessing the ability to interpret exception types, traceback information, and offending lines.
\item \textbf{Error Identification}: Items assessing the ability to locate faults in Python code.
\item \textbf{Error Resolution}: Items assessing the ability to choose an appropriate repair after a fault and cause are known.
\item \textbf{Natural Language Scenarios}: Items requiring participants to infer likely causes of defects from natural-language problem descriptions.
\item \textbf{Miscellaneous}: More complex items combining multiple categories or approximating compact real-world debugging scenarios.
\end{itemize}

In the pilot study, each participant answered 16 items: two randomly sampled items from each theme. Theme order, item order, and answer-option order were randomized.

\subsection{Pilot Procedure}

The proficiency-assessment pilot was implemented in Qualtrics and distributed through convenience sampling. Participants reported their self-assessed Python proficiency, years of experience with Python, years of general programming experience, and their estimate of how many assessment items they had answered correctly.

A total of 78 participants started the survey and 60 completed it, yielding a completion rate of 76.92

\subsection{Pilot Correlations}

Table~\ref{tab:appendix-python-survey-correlations} reports the correlations examined in the pilot. We use these correlations descriptively to understand the relationship among self-reported experience, self-assessed skill, self-estimated accuracy, and actual assessment performance.

\begin{table}[h]
    \centering
    \scriptsize
    \begin{tabularx}{\linewidth}{@{}>{\raggedright\arraybackslash}p{0.20\linewidth}Xcr@{}}
      \toprule
      \textbf{Variable A} & \textbf{Variable B} & \textbf{Corr.} & \textbf{$p$-value} \\
      \midrule
      Python YoE & General programming YoE & $0.55^{\mathrm{P}}$ & 0.00001 \\
      Python YoE & Dreyfus level & $0.52^{\mathrm{S}}$ & 0.00003 \\
      Python YoE & Self-estimated accuracy & $0.24^{\mathrm{S}}$ & 0.00655 \\
      Python YoE & Actual accuracy & $0.28^{\mathrm{P}}$ & 0.01527 \\
      Dreyfus level & Self-estimated accuracy & $0.24^{\mathrm{S}}$ & 0.00655 \\
      General programming YoE & Actual accuracy & $0.06^{\mathrm{P}}$ & 0.62229 \\
      General programming YoE & Self-estimated accuracy & $0.03^{\mathrm{S}}$ & 0.82473 \\
      Actual accuracy & Self-estimated accuracy & $0.35^{\mathrm{S}}$ & 0.00671 \\
      \bottomrule
\end{tabularx}
\caption{Correlations examined in the Python proficiency-assessment pilot. $^{\mathrm{P}}$ denotes Pearson correlation and $^{\mathrm{S}}$ denotes Spearman correlation. Python YoE = self-reported years of experience with Python; General Programming YoE = self-reported years of general programming experience; Dreyfus level = ordinal self-rating of Python experience; accuracy = number of questions answered correctly on a 0--16 scale.}
\label{tab:appendix-python-survey-correlations}
\end{table}

\subsection{Final Item Selection}

To select items for the main study, we calculated item difficulty and point-biserial discrimination. Difficulty is the proportion of participants who answered an item correctly; values closer to 1 indicate easier items. Point-biserial discrimination captures how well an item distinguishes higher-scoring from lower-scoring participants.

We used Classical Test Theory rather than Item Response Theory because the pilot sample was too small for stable item-response estimates. The final eight selected items are shown in Table~\ref{tab:appendix-python-item-diff-disc}.

\begin{table}[h!]
    \centering
    \scriptsize
    \begin{tabularx}{\linewidth}{@{}l>{\centering\arraybackslash}X>{\centering\arraybackslash}Xr@{}}
      \toprule
      \textbf{Question} & \textbf{Difficulty} & \textbf{Discrimination ($r_{pb}$)} & \textbf{$p$-value} \\
      \midrule
      Q4.6  & 0.82 & 0.50 & 0.04224 \\
      Q5.2  & 0.83 & 0.65 & 0.00351 \\
      Q6.5  & 0.84 & 0.60 & 0.00631 \\
      Q7.1  & 0.88 & 0.81 & 0.00008 \\
      Q7.2  & 0.56 & 0.51 & 0.04598 \\
      Q8.2  & 0.78 & 0.40 & 0.10440 \\
      Q9.5  & 0.76 & 0.30 & 0.24669 \\
      Q10.1 & 0.87 & 0.45 & 0.08909 \\
      \bottomrule
\end{tabularx}
\caption{Difficulty, discriminative power, and $p$-value for the final selected items of the Python proficiency assessment.}
\label{tab:appendix-python-item-diff-disc}
\end{table}

Although Q8.2, Q9.5, and Q10.1 did not reach conventional significance thresholds, they were retained because they represented otherwise under-covered themes and had acceptable difficulty and positive discriminative values.

\subsection{Skill Grouping Logic}

In the main study, participants were assigned to lower or higher assessed proficiency using both MCQ score and self-reported Python years of experience. We use the terms lower and higher assessed proficiency rather than novice and expert because the grouping is task-aligned and operational, not a complete measure of programming expertise.

\begin{table}[h]
\centering
\small
\begin{tabularx}{\linewidth}{@{}c>{\centering\arraybackslash}X>{\centering\arraybackslash}X@{}}
\toprule
\textbf{MCQ score} & \textbf{Python experience} & \textbf{Skill group} \\
\midrule
$\geq 6$          & Any                       & Expert \\
$\leq 3$          & Any                       & Novice \\
$[4, 5]$          & $\geq 5$                  & Expert \\
$[4, 5]$          & $< 5$                     & Novice \\
\bottomrule
\end{tabularx}
\caption{Skill grouping logic used in the main study.}
\label{tab:appendix-skill-level-logic}
\end{table}

\section{Debugging Task Selection}
\label{appendix:debugging-tasks}

This appendix provides additional detail on how the debugging tasks were generated, screened, piloted, and selected for the main study.

\subsection{Candidate Generation}

We generated eight candidate buggy Python snippets targeting common error families, including \texttt{NameError}, \texttt{TypeError}, and \texttt{SyntaxError}. Each candidate was required to satisfy the following constraints:

\begin{itemize}
\item \textbf{Self-contained}: The program could be run as a standalone Python script.
\item \textbf{Standard library only}: The snippet did not require external packages.
\item \textbf{Moderate length}: The snippet was limited to at most 60 lines.
\item \textbf{Moderate structure}: The snippet included several functions or class methods.
\item \textbf{Single primary fault}: The snippet was designed to trigger one primary error.
\item \textbf{Hidden-test compatible}: The intended repair could be evaluated using a semantic test suite.
\end{itemize}

Each candidate was manually checked to ensure that it triggered the intended error and did not contain multiple simultaneous faults. When candidates violated these constraints, they were corrected or regenerated.

\subsection{Task-Selection Pilot}

Participants in the task-selection pilot rated each candidate snippet and its standard Python interpreter message on four dimensions: code difficulty, fix difficulty, error-message mental demand, and error-message usefulness. Table~\ref{tab:appendix-task-selection-items} lists the exact survey items.

\begin{table}[h!]
\centering
\begin{tabular}{p{0.25\linewidth} p{0.65\linewidth}}
\toprule
\textbf{Construct assessed} & \textbf{Survey item} \\
\midrule
Code difficulty & ``\textit{This code snippet is difficult to understand.}'' \\
Fix difficulty & ``\textit{I would find it challenging to resolve the issue in this code snippet.}'' \\
Error mental demand & ``\textit{Reading this error message feels mentally demanding.}'' \\
Error usefulness & ``\textit{This error message is useful for identifying the problem.}'' \\
\bottomrule
\end{tabular}
\caption{Mapping between the pilot study's survey dimensions and their corresponding items.}
\label{tab:appendix-task-selection-items}
\end{table}

The task-selection pilot included 20 valid responses after excluding one participant who failed an attention check.

\subsection{Selection Criteria}

The final four snippets were selected to provide coverage across error types while remaining appropriate for the target participant population. We prioritized candidates that met the following criteria:

\begin{itemize}
\item moderate perceived code difficulty;
\item moderate perceived fix difficulty;
\item variation in error type and fault structure;
\item interpretable standard interpreter message;
\item absence of multiple primary faults;
\item compatibility with hidden semantic tests.
\end{itemize}

\subsection{Selected Snippets and Standard Error Messages}

The selected snippets are provided below. Each snippet was shown to participants together with one assigned error-message style in the main study.

\subsubsection{Snippet A}
\label{appendix:snippet-a}

\begin{lstlisting}[style=mypython, caption={Selected debugging task: Snippet A.}, label={lst:appendix-snippet-a}, captionpos=b]
% Insert Snippet A code here.
\end{lstlisting}

\begin{lstlisting}[basicstyle=\ttfamily\footnotesize, breaklines=true, caption={Standard Python interpreter message for Snippet A.}, label={lst:appendix-snippet-a-error}, captionpos=b]
% Insert standard interpreter error message for Snippet A here.
\end{lstlisting}

\subsubsection{Snippet B}
\label{appendix:snippet-b}

\begin{lstlisting}[style=mypython, caption={Selected debugging task: Snippet B.}, label={lst:appendix-snippet-b}, captionpos=b]
import random

class UserData:
"""Represents user data with a name and a list of scores."""

```
def __init__(self, name, scores):
    self.name = name
    self.scores = scores

def top_score(self):
    """Returns the highest score."""
    return maximum(self.scores) if self.scores else 0

def add_score(self, score):
    self.scores.append(score)
```

def summarize_scores(users):
return {u.name: u.top_score() for u in users}

if **name** == '**main**':
"""Main routine to generate user data, summarize scores, and print the results."""
users = [UserData(f"user_{i + 1}", [random.randint(0, 100) for _ in range(random.randint(2, 5))]) for i in range(4)]
summary = summarize_scores(users)
for name, score in summary.items():
print(f"{name}: {score:.2f}")
\end{lstlisting}

\begin{lstlisting}[basicstyle=\ttfamily\footnotesize, breaklines=true, caption={Standard Python interpreter message for Snippet B.}, label={lst:appendix-snippet-b-error}, captionpos=b]
% Insert standard interpreter error message for Snippet B here.
\end{lstlisting}

\subsubsection{Snippet C}
\label{appendix:snippet-c}

\begin{lstlisting}[style=mypython, caption={Selected debugging task: Snippet C.}, label={lst:appendix-snippet-c}, captionpos=b]
% Insert Snippet C code here.
\end{lstlisting}

\begin{lstlisting}[basicstyle=\ttfamily\footnotesize, breaklines=true, caption={Standard Python interpreter message for Snippet C.}, label={lst:appendix-snippet-c-error}, captionpos=b]
% Insert standard interpreter error message for Snippet C here.
\end{lstlisting}

\subsubsection{Snippet D}
\label{appendix:snippet-d}

\begin{lstlisting}[style=mypython, caption={Selected debugging task: Snippet D.}, label={lst:appendix-snippet-d}, captionpos=b]
% Insert Snippet D code here.
\end{lstlisting}

\begin{lstlisting}[basicstyle=\ttfamily\footnotesize, breaklines=true, caption={Standard Python interpreter message for Snippet D.}, label={lst:appendix-snippet-d-error}, captionpos=b]
% Insert standard interpreter error message for Snippet D here.
\end{lstlisting}

\section{LLM Prompt Templates and Generation Settings}
\label{appendix:prompts}

This appendix provides the full prompt templates and generation settings used to create the LLM-generated programming error messages.

\subsection{Generation Model}

All participant-facing rewritten messages were generated using \texttt{llama-3.1-8B-Instruct}. The model was selected because it could be self-hosted, followed instructions reliably in pilot testing, supported sufficient context for the selected snippets and tracebacks, and allowed deterministic generation. The candidate buggy snippets used during task selection were generated separately using \texttt{GPT-4o-mini-high}; however, all messages shown to participants in the main study were generated using \texttt{llama-3.1-8B-Instruct}.

\subsection{Generation Settings}

We used zero-shot prompting and set temperature to 0. This setting was chosen to improve consistency and reproducibility across generated messages. In pilot testing, one-shot and few-shot prompts biased outputs toward the examples and reduced generality.

Each prompt included:

\begin{itemize}
\item the full code snippet;
\item the standard Python interpreter error message;
\item line-numbered code context;
\item instructions to identify the cause and relevant line from the code, rather than copying the traceback line number alone.
\end{itemize}

\subsection{Pragmatic Prompt Template}

\begin{tcolorbox}[title=Pragmatic prompt template,
label=box:appendix-pragmatic,
colback=gray!10,
colframe=black!40,
fonttitle=\bfseries,
breakable=true]
\begin{lstlisting}[basicstyle=\ttfamily\scriptsize,breaklines=true,breakatwhitespace=true,columns=fullflexible,keepspaces=true,showstringspaces=false]
INSTRUCTION:
You are an assistant helping a Python programmer by explaining the error, in a clear and
concise way.

CONTEXT:
The programmer's code and error message are below:
--------------------------------------------------

CODE:

```python
{code}
```

ERROR MESSAGE:

```
{error}
```

---

TASK:

1. Identify the cause of the error and the relevant line number from the code snippet.
   DO NOT FETCH THE LINE NUMBER FROM THE ERROR MESSAGE.
2. Write exactly one paragraph (around 20-25 words or less) that:

   * Begins with "**<ExceptionType>** at **line <line>**:"
   * Briefly states the cause and hints at a fix.
   * Focuses on providing helpful actionable insights, without directly giving the
     corrected code.

FORMAT:
You may use markdown for emphasis, but do NOT include lists or code fences.

NOW WRITE YOUR RESPONSE:
\end{lstlisting}
\end{tcolorbox}

\subsection{Contingent Prompt Template}

\begin{tcolorbox}[title=Contingent prompt template,
label=box:appendix-contingent,
colback=gray!10,
colframe=black!40,
fonttitle=\bfseries,
breakable=true]
\begin{lstlisting}[basicstyle=\ttfamily\scriptsize,breaklines=true,breakatwhitespace=true,columns=fullflexible,keepspaces=true,showstringspaces=false]
INSTRUCTION:
Guide the Python programmer to resolve their error in 3-5 supportive sentences. Focus on
actionable steps and encouragement, while avoiding an authoritative tone.

CONTEXT:
The programmer's code and error message are below:
--------------------------------------------------

CODE:

```python
{code}
```

ERROR MESSAGE:

```
{error}
```

---

TASK:

1. Identify the cause of the error and the relevant line number from the code snippet.
   DO NOT FETCH THE LINE NUMBER FROM THE ERROR MESSAGE.
2. Then write 3-5 sentences that:

   * Begin your response with "**<ExceptionType>** at **line <line>**:" on the first line.
   * Then confirm the likely intent or goal (e.g. "Did you mean to ...?") - this is the
     claim, showing you understand what the programmer was trying to do.
   * Only if useful, mention whether this error is common or if there are exceptions.
   * Only if relevant, note whether there are situations where the error might occur
     differently.

FORMAT:
You may use markdown for emphasis, but do NOT include lists or code fences.

NOW WRITE YOUR RESPONSE:
\end{lstlisting}
\end{tcolorbox}

\section{Main Study Details}
\label{appendix:main-study-details}

This appendix provides additional details on recruitment, study workflow, interface design, quality control, and survey items for the main Prolific experiment.

\subsection{Recruitment and Eligibility}

Participants were recruited through Prolific. Eligibility criteria required participants to be at least 18 years old, fluent in English, have prior Python experience according to the Prolific screener, have an approval rate above 90

\subsection{Platform}

The study was deployed through a custom web application. Prolific was used for recruitment, eligibility screening, and compensation management; the experimental tasks were administered on our platform. Participants arrived from Prolific with a unique Prolific ID in the URL, allowing us to enforce one completion per participant and align task records.

The application used a Next.js/React frontend, a FastAPI backend, and PostgreSQL for data storage. The debugging task was presented in an embedded code editor with Python syntax highlighting and non-AI rule-based inline completion. The system executed submissions against hidden semantic tests to determine whether the targeted bug had been repaired.

\subsection{Participant Workflow}

After providing consent, participants completed the following stages:

\begin{enumerate}
\item \textbf{Instructions and sandbox}: Participants reviewed task instructions and interacted with a sandbox version of the interface.
\item \textbf{Skill assessment}: Participants reported years of Python experience and answered the eight selected MCQ items.
\item \textbf{Debugging task}: Participants received one calibrated buggy Python snippet and one assigned error-message style.
\item \textbf{Repair attempts}: Participants had up to three attempts to fix the code. After each failed attempt, the editor reset to the original code.
\item \textbf{Post-task survey}: Participants rated the error message they received on readability, cognitive load, and tone.
\end{enumerate}

\subsection{Assignment and Balancing}

Participants were assigned to one of three message styles: standard, pragmatic, or contingent. Assignment was constrained to maintain balance across message styles and snippets. Each participant saw exactly one snippet and one message style.

\subsection{Quality Control}

The application logged integrity and quality-control signals, including copy, paste, cut, browser focus changes, tab switching, completion time, and repeated submissions. Participants showing clear evidence of low-effort behavior, random typing, repeated unchanged submissions, or likely external assistance were manually reviewed and excluded according to the study criteria disclosed to participants.

\subsection{Outcome Measures}

The main study measured both objective debugging outcomes and subjective message evaluations.

\paragraph{Objective outcomes.}
Objective outcomes included:

\begin{itemize}
\item \textbf{Fix rate}: whether the participant successfully repaired the code within three attempts.
\item \textbf{Fix@k}: whether the participant first succeeded on attempt 1, 2, or 3.
\item \textbf{Time-to-fix}: elapsed time from task start until the first correct submission, for participants who successfully repaired the snippet.
\item \textbf{Number of attempts}: the number of submissions made before success or task termination.
\end{itemize}

\paragraph{Subjective outcomes.}
After the debugging task, participants rated the message they received using 5-point Likert items covering readability, cognitive load, and perceived tone. The full items are shown in Table~\ref{tab:appendix-survey-feedback-items}.

\begin{table}[htbp!]
\centering
\renewcommand{\arraystretch}{1.2}
\newcolumntype{L}[1]{>{\raggedright\arraybackslash}p{#1}}
\begin{tabularx}{\linewidth}{L{0.22\linewidth} L{0.43\linewidth} L{0.27\linewidth}}
\toprule
\textbf{Metric} & \textbf{Question} & \textbf{Scale (1-5)} \\
\midrule
\multicolumn{3}{l}{\textbf{Readability}} \\
\midrule
Length &
``\textit{Is the message expressed using more words than needed?}'' &
Succinct - Verbose \\
Jargon &
``\textit{Does the message contain jargon and technical terms?}'' &
Less - More \\
Sentence Structure &
``\textit{How clear is the sentence structure of the message?}'' &
Clear - Unclear \\
Vocabulary &
``\textit{How complex is the vocabulary used?}'' &
Simple - Complex \\
\midrule
\multicolumn{3}{l}{\textbf{Cognitive load}} \\
\midrule
Intrinsic Load &
``\textit{This error message was inherently difficult to understand.}'' &
Disagree - Agree \\
Extraneous Load &
``\textit{The wording or formatting of this error message wasted mental effort.}'' &
Disagree - Agree \\
Germane Load &
``\textit{This error message helped me recognize the underlying error concept.}'' &
Disagree - Agree \\
\midrule
\multicolumn{3}{l}{\textbf{Authoritativeness}} \\
\midrule
Tone (respectfulness) &
``\textit{How respectful (i.e., reader-centered) is the tone of the error message?}'' &
Disrespectful - Respectful \\
\bottomrule
\end{tabularx}
\caption{Survey items for readability, cognitive load, and authoritativeness, with response scales.}
\label{tab:appendix-survey-feedback-items}
\end{table}

\subsection{Participant Allocation}

The final dataset included 103 valid participants after quality control. Table~\ref{tab:allocation} shows allocation by message style and assessed proficiency group. Table~\ref{tab:appendix-allocation-snippet-style} shows allocation by snippet and message style.

\begin{table}[h!]
\centering
\begin{tabular}{lcccc}
\toprule
 & \textbf{Standard} & \textbf{Pragmatic} & \textbf{Contingent} & \textbf{Total} \\
\midrule
\textbf{Novice} & 13 & 13 & 12 & 38 \\
\textbf{Expert} & 22 & 22 & 21 & 63 \\
\midrule
\textbf{Total} & 35 & 35 & 33 & 103 \\
\bottomrule
\end{tabular}
\caption{Participant allocation by message style and skill level.}
\label{tab:allocation}
\end{table}

\begin{table}[h!]
\centering
\begin{tabular}{lccccc}
\toprule
 & \textbf{A} & \textbf{B} & \textbf{C} & \textbf{D} & \textbf{Total} \\
\midrule
\textbf{Standard}   & 9 & 8 & 9 & 9 & 35 \\
\textbf{Pragmatic}  & 9 & 9 & 8 & 9 & 35 \\
\textbf{Contingent} & 8 & 9 & 8 & 8 & 33 \\
\midrule
\textbf{Total}      & 26 & 26 & 25 & 26 & 103 \\
\bottomrule
\end{tabular}
\caption{Participant allocation by snippet (A-D) within each message style.}
\label{tab:appendix-allocation-snippet-style}
\end{table}

\begin{figure}[h!]
    \centering

    \begin{subfigure}[b]{0.48\linewidth}
        \centering
        \includegraphics[width=\linewidth]{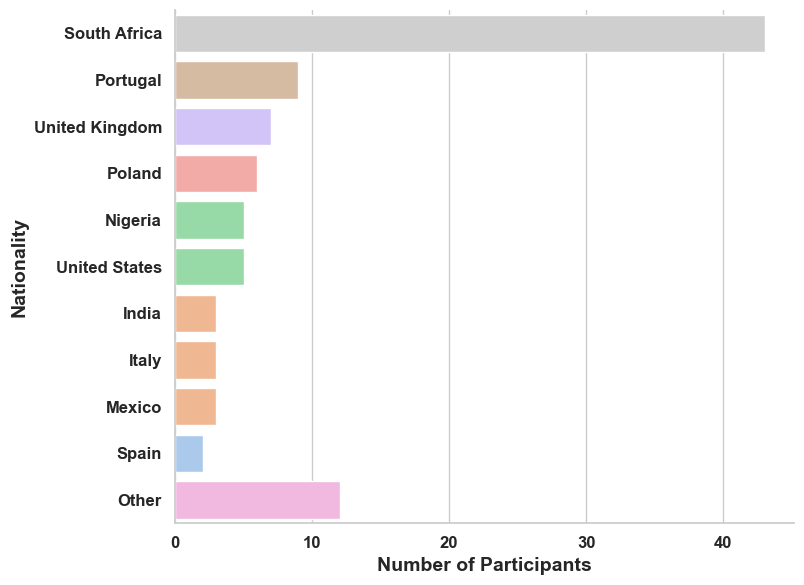}
        \caption{Participants' nationality distribution.}
        \label{fig:prolific-nationality-distribution}
    \end{subfigure}
    \hfill
    \begin{subfigure}[b]{0.48\linewidth}
        \centering
        \includegraphics[width=\linewidth]{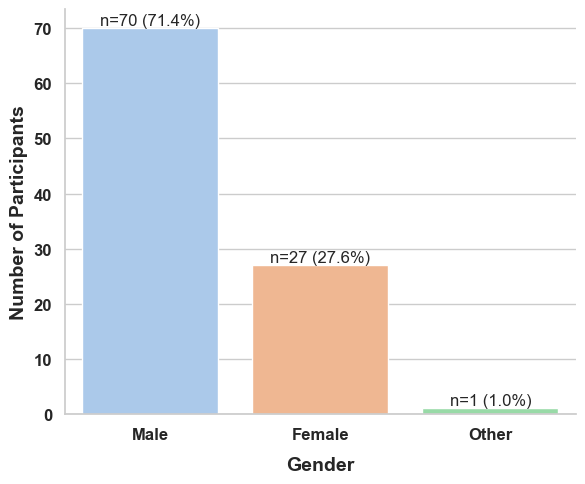}
        \caption{Participants' gender distribution.}
        \label{fig:prolific-gender-distribution}
    \end{subfigure}

    \par\medskip

    \begin{subfigure}[b]{0.6\linewidth}
        \centering
        \includegraphics[width=\linewidth]{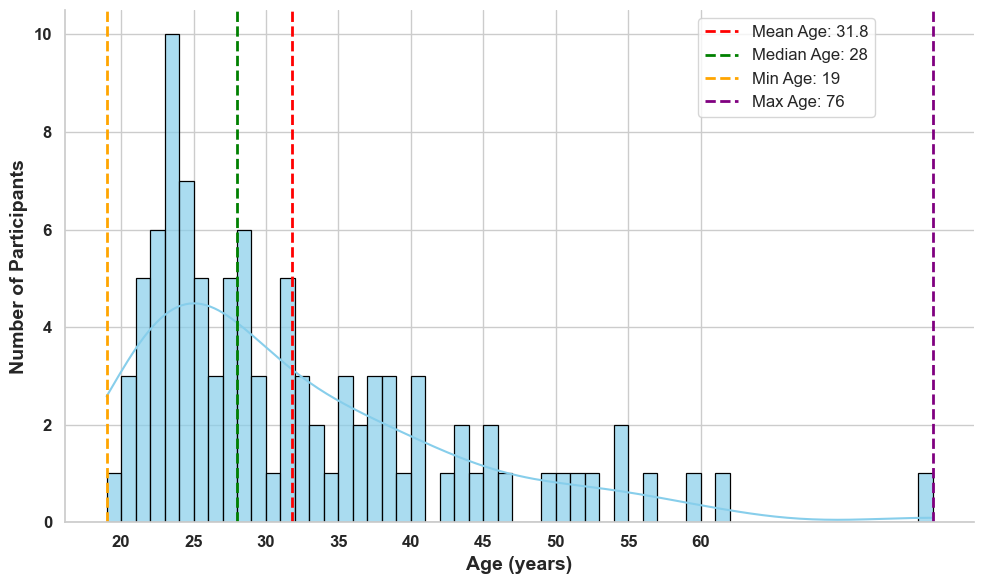}
        \caption{Participants' age distribution.}
        \label{fig:prolific-age-distribution}
    \end{subfigure}

    \caption{Descriptive statistics for Prolific participants' demographics.}
    \label{fig:prolific-descriptive-statistics}
\end{figure}

%% file: appendix/appendix_a.tex
\newpage

\section{Python Skill Level Assessment Qualtrics Questionnaire and Details}
\label{appendix:qualtrics-full-question-list}

This appendix collects material that was used for, or is directly relevant to, the Python Skill Level Assessment that informed our pilot study, which we launched on Qualtrics\footnote{\url{https://www.qualtrics.com/}}. It brings together the elements that shaped the assessment and the specific items that we later relied on in the main study\footnote{Readers ought to remember that within the main contents of this paper, Section \ref{sec:python-assessment} describes the motivation for creating the assessment and the approach we followed during the pilot phase. It explains why a dedicated instrument was needed and how we constructed and refined it.}.

Table \ref{tab:experience-levels-dreyfus} presents the self rating question for Python experience that we used in the pilot study. The question uses six levels that are informed by the Dreyfus model of skill acquisition. The table shows the exact wording of the question and the response options so that the reader can see how participants reported their level of experience.

\begin{figure}[h!]
  \centering

  \begin{subfigure}[b]{0.48\linewidth}
    \centering
    \includegraphics[width=\linewidth]{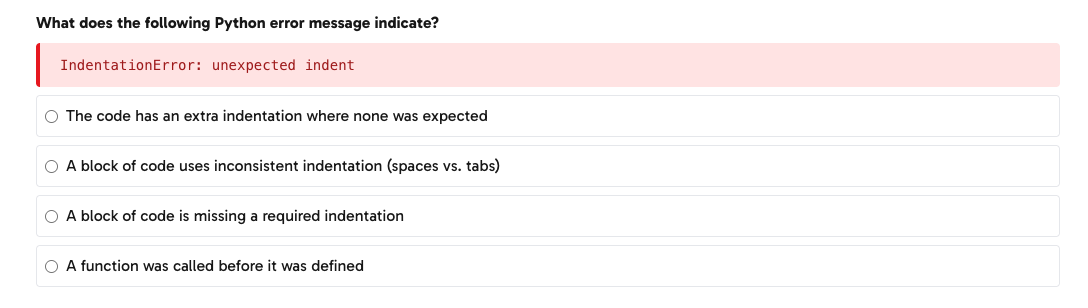}
    \caption{Question ID Q4.6 of the Python Qualtrics Skill-Assessment Survey.}
    \label{fig:qid-4-6}
  \end{subfigure}
  \hfill
  \begin{subfigure}[b]{0.48\linewidth}
    \centering
    \includegraphics[width=\linewidth]{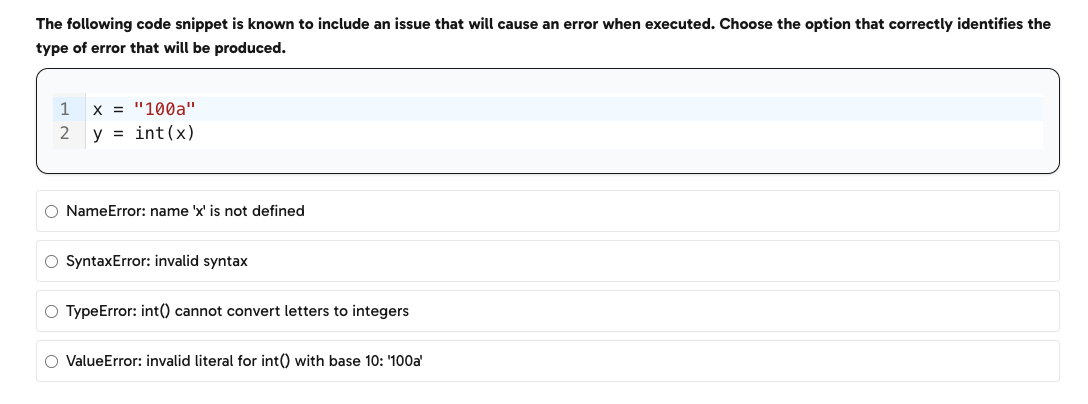}
    \caption{Question ID Q5.2 of the Python Qualtrics Skill-Assessment Survey.}
    \label{fig:qid-5-2}
  \end{subfigure}

  \vspace{0.6em}

  \begin{subfigure}[b]{0.48\linewidth}
    \centering
    \includegraphics[width=\linewidth]{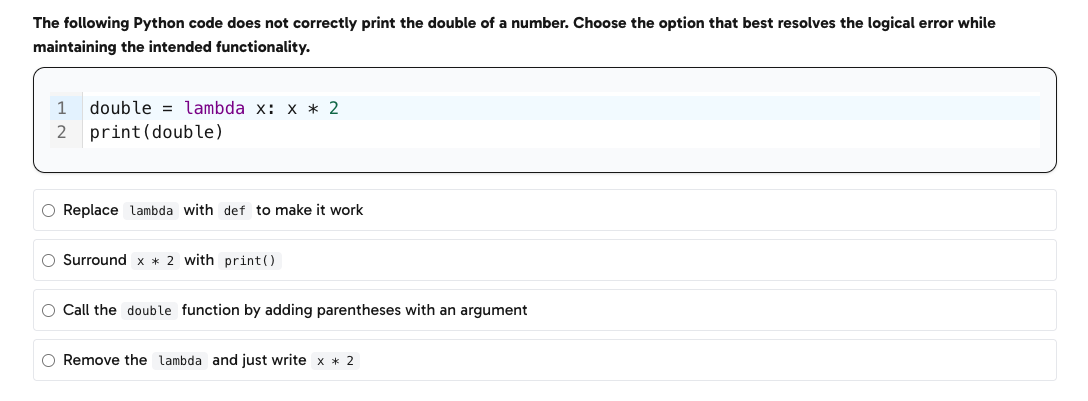}
    \caption{Question ID Q6.5 of the Python Qualtrics Skill-Assessment Survey.}
    \label{fig:qid-6-5}
  \end{subfigure}
  \hfill
  \begin{subfigure}[b]{0.48\linewidth}
    \centering
    \includegraphics[width=\linewidth]{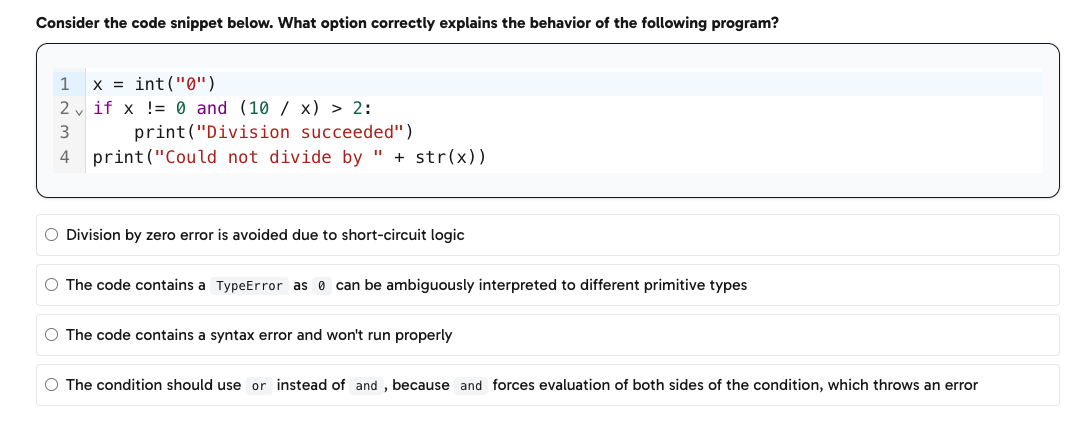}
    \caption{Question ID Q7.1 of the Python Qualtrics Skill-Assessment Survey.}
    \label{fig:qid-7-1}
  \end{subfigure}

  \vspace{0.6em}

  \begin{subfigure}[b]{0.48\linewidth}
    \centering
    \includegraphics[width=\linewidth]{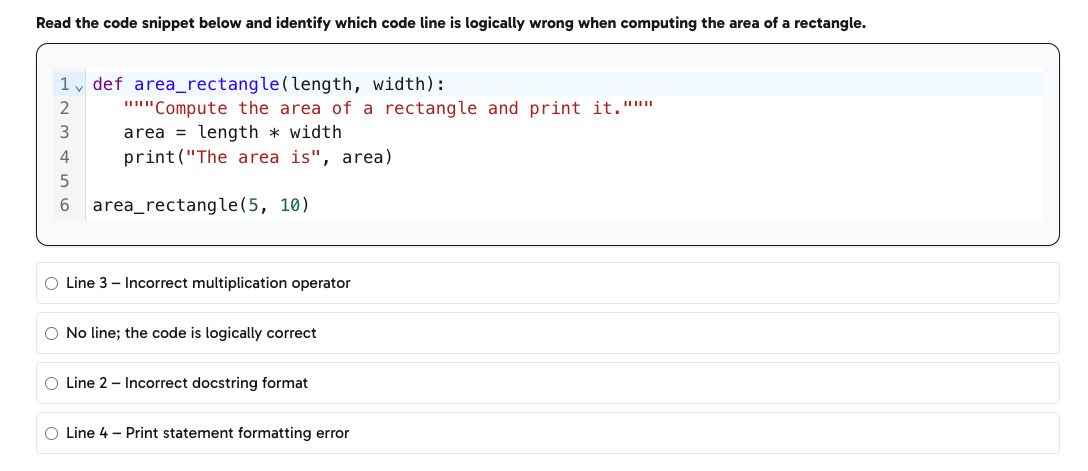}
    \caption{Question ID Q7.2 of the Python Qualtrics Skill-Assessment Survey.}
    \label{fig:qid-7-2}
  \end{subfigure}
  \hfill
  \begin{subfigure}[b]{0.48\linewidth}
    \centering
    \includegraphics[width=\linewidth]{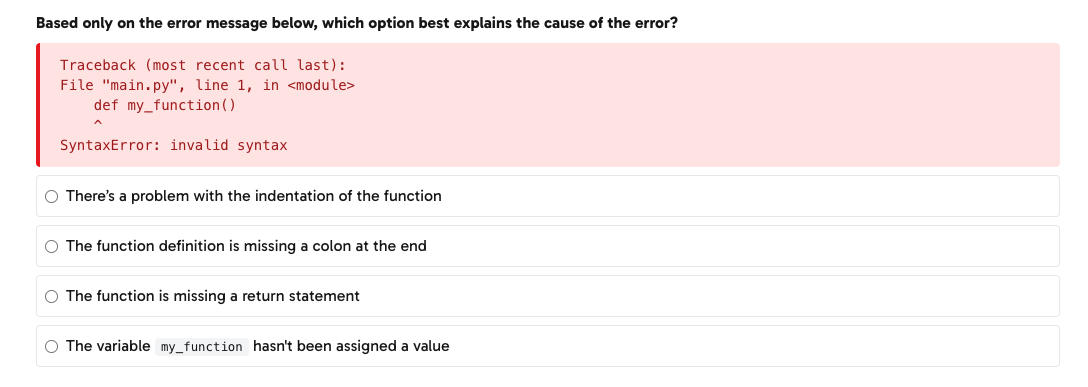}
    \caption{Question ID Q8.2 of the Python Qualtrics Skill-Assessment Survey.}
    \label{fig:qid-8-2}
  \end{subfigure}

  \vspace{0.6em}

  \vspace{0.6em}
  \begin{subfigure}[b]{0.48\linewidth}
    \centering
    \includegraphics[width=\linewidth]{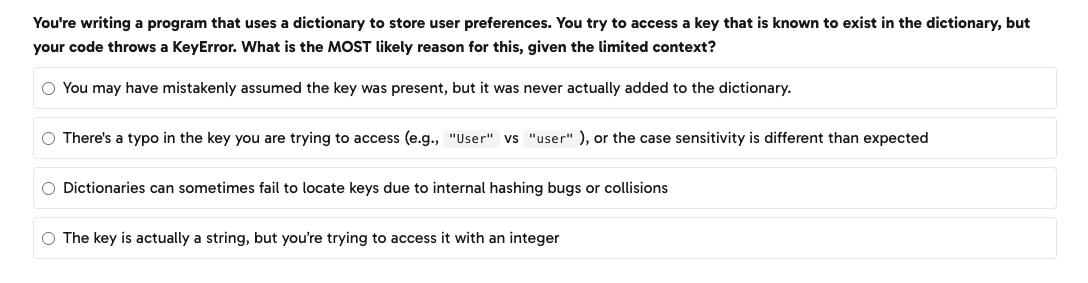}
    \caption{Question ID Q9.5 of the Python Qualtrics Skill-Assessment Survey.}
    \label{fig:qid-9-5}
  \end{subfigure}
  \hfill
  \begin{subfigure}[b]{0.48\linewidth}
    \centering
    \includegraphics[width=\linewidth]{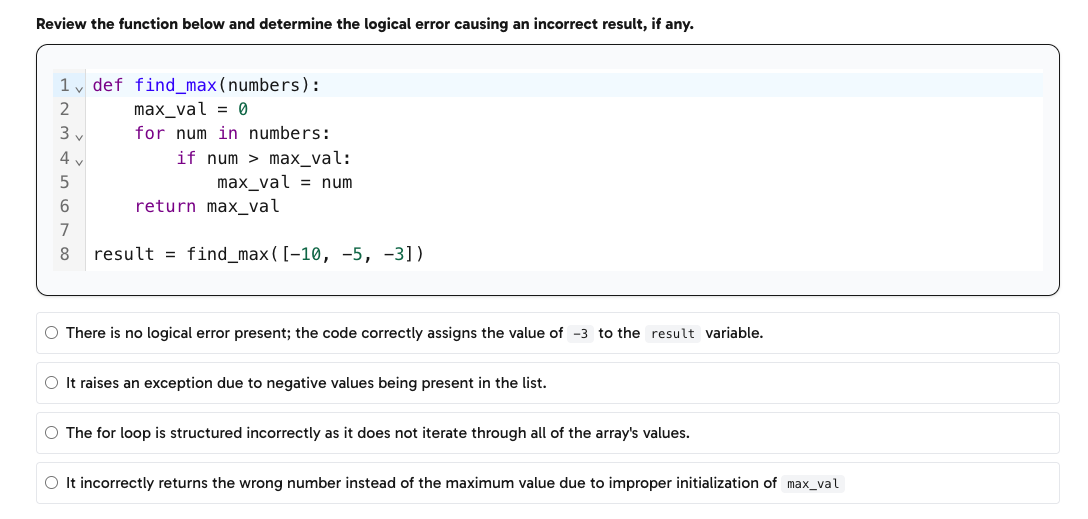}
    \caption{Question ID Q10.1 of the Python Qualtrics Skill-Assessment Survey.}
    \label{fig:qid-10-1}
  \end{subfigure}

  \caption{Final set of survey items that most effectively discriminated between novices and experts in the Python Skill-Assessment Survey. The images reproduce the question options and interface, as presented in the Prolific web application during the main study.}
  \label{fig:qualtrics-selected-items}
\end{figure}

Figure \ref{fig:qualtrics-selected-items} shows the eight multiple choice questions that we selected from an initial pool of 56 items\footnote{For brevity, we do not include all 56 pilot questions here. The full set, along with additional plots, and the analysis scripts and notebooks used to select the final eight, is available in our supplementary material.}. These eight items were identified as the most capable of distinguishing novices from experts in the context of debugging and understanding error messages. As explained in Section \ref{sec:python-assessment}, this set of eight items was then carried forward and used in the main Prolific study.

\begin{table}[h]
  \centering
  \begin{tabularx}{\linewidth}{c X}
    \toprule
    \textbf{Numerical Level} & \textbf{Experience level statement} \\
    \midrule
    1 & I have no previous experience in Python. \\
    2 & Focusing on syntax \& basics, relying on tutorials, lacking real-world project experience. \\
    3 & Practical experience, small projects or assignments. Grasping basic concepts, and troubleshooting independently. \\
    4 & Experience with projects. Able to independently plan and execute tasks, proficient with Python's libraries. \\
    5 & Experience in complex projects. Deep understanding, ability to consciously resolve difficult problems \\
    6 & Significant experience in larger complex programs. Solves complex problems effortlessly and subconsciously \\
    \bottomrule
  \end{tabularx}
  \caption{Six-level self-rating of Python experience used in the study, informed by the Dreyfus skill-acquisition model.}
  \label{tab:experience-levels-dreyfus}
\end{table}

%% file: appendix/appendix_b.tex
\newpage

\section{Fantastic Four Code Snippets and Error Messages}
\label{appendix:fantastic four code snippets}

This appendix presents the four buggy Python snippets used in the Prolific study, with their standard interpreter errors. Sections \ref{sec:snippet-a-appendix}, \ref{sec:snippet-b-appendix}, \ref{sec:snippet-c-appendix}, \ref{sec:snippet-d-appendix} show each code snippet followed by its error message.

\subsection{Snippet A: \texttt{SyntaxError: unterminated string literal}}
\label{sec:snippet-a-appendix}

\bigskip

\begin{lstlisting}[style=mypython, caption={One of the chosen four buggy Python code snippet, which when executed as a standalone file, will trigger a \texttt{SyntaxError}, due to the unclosed triple-quoted string literal on line 17.}, label={lst:fantastic-four-snippet-a}, captionpos=b]
class BookShelf:
    def __init__(self, log):
        self.log = log

    def preview(self):
        """
        Shows first two book titles, if any.
        """
        if not os.path.exists(self.log):
            return "No log found."
        with open(self.log) as f:
            lines = f.readlines()
        preview = "".join(lines[:2])
        return f"""Preview:\n{preview}""

    def summary(self):
        """
        Gives a summary of the log.
        """
        total = count_books(self.log)
        return f"Books logged: {total}"

def add_book(log, title):
    """
    Adds a book entry with timestamp.
    """
    with open(log, 'a') as f:
        f.write(f"{datetime.now().isoformat()} - {title}\n")

def count_books(log):
    """
    Counts books in the log.
    """
    if not os.path.exists(log):
        return 0
    with open(log) as f:
        return sum(1 for _ in f)
\end{lstlisting}

\begin{figure}[h]
  \centering
  \includegraphics[width=0.55\linewidth]{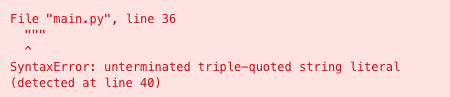}
  \caption{Standard Python interpreter message displayed when running the code in Listing \ref{lst:fantastic-four-snippet-a}.}
  \label{fig:snippet-a-error-message}
\end{figure}

\newpage

\subsection{Snippet B: \texttt{NameError: name not defined}}
\label{sec:snippet-b-appendix}

\bigskip

\begin{lstlisting}[style=mypython, caption={One of the chosen four buggy Python code snippet, which when executed as a standalone file, will trigger a \texttt{NameError} on line 13 of the code listing, due to the undefined variable \texttt{maximum}.}, label={lst:fantastic-four-snippet-b}, captionpos=b]
import random

class UserData:
    """Represents user data with a name and a list of scores."""

    def __init__(self, name, scores):
        self.name = name
        self.scores = scores

    def top_score(self):
        """Returns the highest score."""
        return maximum(self.scores) if self.scores else 0

    def add_score(self, score):
        self.scores.append(score)

def summarize_scores(users):
    return {u.name: u.top_score() for u in users}

if __name__ == '__main__':
    """Main routine to generate user data, summarize scores, and print the results."""
    users = [UserData(f"user_{i + 1}", [random.randint(0, 100) for _ in range(random.randint(2, 5))]) for i in range(4)]
    summary = summarize_scores(users)
    for name, score in summary.items():
        print(f"{name}: {score:.2f}")
\end{lstlisting}

\begin{figure}[h]
  \centering
  \includegraphics[width=0.55\linewidth]{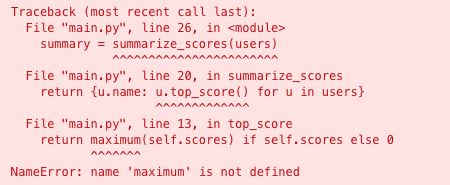}
  \caption{Standard Python interpreter message displayed when running the code in Listing \ref{lst:fantastic-four-snippet-b}.}
  \label{fig:snippet-b-error-message}
\end{figure}

\newpage

\subsection{Snippet C: \texttt{TypeError: unsupported operand types}}
\label{sec:snippet-c-appendix}

\bigskip

\begin{lstlisting}[style=mypython, caption={One of the chosen four buggy Python code snippet, which when executed as a standalone file, will trigger a \texttt{TypeError} on line 21 of the code listing, due to adding both \texttt{@classmethod} and \texttt{@staticmethod} on the same line.}, label={lst:fantastic-four-snippet-c}, captionpos=b]
import random

def generate_scores(n):
    """Generate n random test scores."""
    return [random.randint(0, 100) for _ in range(n)]

def average(scores):
    """Compute the average score."""
    if not scores:
        return 0
    return sum(scores) / len(scores)

def filter_passing(scores, threshold=60):
    """Return scores that are passing."""
    return [s for s in scores if s >= threshold]

class ScoreReport:
    def __init__(self, scores):
        self.scores = scores

    @classmethod @staticmethod
    def describe():
        """Describe the scoring system."""
        return "Scores range from 0 to 100."

    def passing_percentage(self):
        """Return the percentage of passing scores."""
        passing = filter_passing(self.scores)
        return 100 * len(passing) / len(self.scores) if self.scores else 0

    def report(self):
        """Return a formatted report."""
        avg = average(self.scores)
        pct = self.passing_percentage()
        desc = self.describe()
        return f"{desc}\nAverage: {avg:.1f}\nPassing: {pct:.1f}%"

def main():
    scores = generate_scores(12)
    report = ScoreReport(scores)
    print(report.report())

if __name__ == "__main__":
    main()
\end{lstlisting}

\bigskip

\begin{figure}[h]
  \centering
  \includegraphics[width=0.5\linewidth]{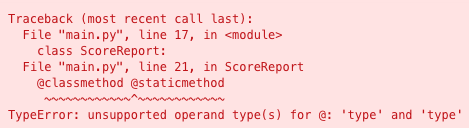}
  \caption{Standard Python interpreter message displayed when running the code in Listing \ref{lst:fantastic-four-snippet-c}.}
  \label{fig:snippet-c-error-message}
\end{figure}

\newpage

\subsection{Snippet D: \texttt{TypeError: object is not subscriptable}}
\label{sec:snippet-d-appendix}

\bigskip

\begin{lstlisting}[style=mypython, caption={One of the chosen four buggy Python code snippet, which when executed as a standalone file, will trigger a \texttt{TypeError} on line 39 of the code listing, due to using the wrong syntax for accessing a list's elements.}, label={lst:fantastic-four-snippet-d}, captionpos=b]
import math

def normalize(vec):
    norm = math.sqrt(sum(x ** 2 for x in vec))
    return [x / norm for x in vec] if norm else vec

def dot(a, b):
    return sum(x * y for x, y in zip(a, b))

def cosine(a, b):
    if len(a) != len(b): raise ValueError("Vectors must be of the same length")
    return dot(normalize(a), normalize(b))

def fixed_vectors():
    """Returns a fixed set of vectors for testing purposes."""
    return [
        [1.0, 2.0, 3.0],
        [2.0, 0.0, 1.0],
        [-1.0, 1.0, 0.0],
        [0.5, -2.0, 2.0]
    ]

def most_similar_pair(vectors):
    """Finds the most similar pair of vectors based on cosine similarity."""
    max_sim = -2
    pair = (0, 1)
    for i in range(len(vectors)):
        for j in range(i + 1, len(vectors)):
            sim = cosine(vectors[i], vectors[j])
            if sim > max_sim:
                max_sim = sim
                pair = (i, j)
    return pair

def main():
    vs = fixed_vectors()
    print("Most similar pair:", most_similar_pair(vs))
    for i in range(len(vs)):
        print("Vector", i, ":", vs.__getitem__[i])

if __name__ == "__main__":
    main()
\end{lstlisting}

\bigskip

\begin{figure}[h]
  \centering
  \includegraphics[width=0.55\linewidth]{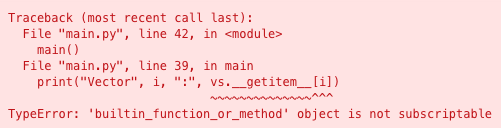}
  \caption{Standard Python interpreter message displayed when running the code in Listing \ref{lst:fantastic-four-snippet-d}.}
  \label{fig:snippet-d-error-message}
\end{figure}

%% file: appendix/appendix_c.tex
\newpage 
\section{Prolific Full-Stack Web Application Design Details}
\label{appendix:prolific-web-application-design-details}

This appendix describes the full-stack platform we deployed on Prolific for the main study and explains the key design decisions behind it. We coordinated services with Docker Compose\footnote{\url{https://docs.docker.com/compose/}} to keep components modular and easy to start, stop, and update. Containerization also allowed us to deploy everything on a Linux virtual machine with only the \texttt{docker} binary installed. This reduced configuration drift over time as we developed the application, and improved reproducibility, which was an important requirement for our study.

\subsection{System Architecture}

\begin{figure}[h!]
  \centering
  \includegraphics[width=0.75\linewidth]{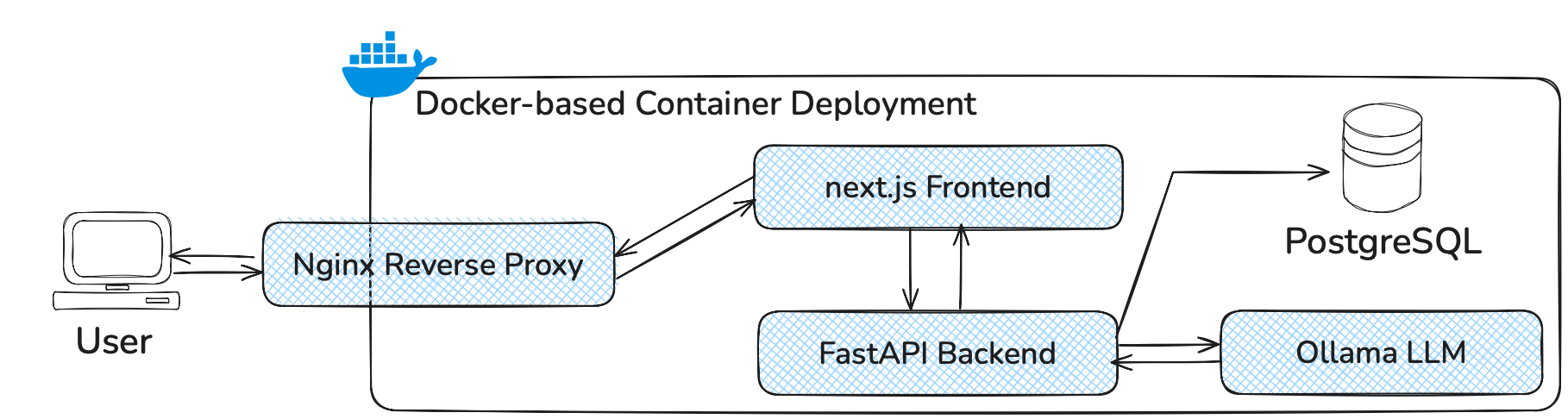}
  \caption{System architecture of the Prolific study full-stack application: containerized frontend, backend, database, local LLM inference, and reverse proxy orchestrated with Docker Compose.}
  \label{fig:prolific-web-app-system-design}
\end{figure}

The platform consists of a web frontend, a Python backend, a relational database, a local LLM inference service, and a reverse proxy. Figure~\ref{fig:prolific-web-app-system-design} shows the high-level architecture and how these components interact within a containerized environment.

The frontend was implemented with Next.js, React, Typescript, and TailwindCSS. We chose this stack mostly due to familiarity and because it was straightforward to build reusable components, such as forms, multi-step submission flows, and interactive elements for task execution and feedback collection.

The backend was built with FastAPI\footnote{\url{https://fastapi.tiangolo.com/}} in Python. FastAPI offers strong type support, automatic API documentation, and efficient asynchronous I/O. These features suited our needs for handling user-provided code execution and returning LLM-rephrased error messages. Keeping the backend in Python also simplified integration with our evaluation logic and the LLM client used for inference.

We stored participant data, submissions, feedback, and event logs in PostgreSQL. PostgreSQL provides mature tooling, transactional guarantees, and expressive queries. A relational schema with ACID properties helped maintain integrity across multi-step tasks, for example when linking code submissions to feedback and timing events. Given our relatively modest scale, any other relational or non-relational database would have been equally suitable for our requirements, yet PostgreSQL aligned best with our planned analysis workflow.

Local LLM inference was provided by Ollama. Running inference on a self-hosted service gave us control over data handling and latency while not incurring additional costs by using an external API, such as OpenAI's API. The service loaded the model used for error-message rephrasing. Our application can technically support any other open-source models available on Ollama within the available memory and latency budget. However, as described in Section~\ref{sec:message-conditions}, we selected \texttt{llama-3.1-8B-Instruct}.

Finally, Nginx\footnote{\url{https://nginx.org/}} served as the reverse proxy. It routed requests to the frontend and backend and cached static assets to improve responsiveness during data collection. It also restricted public access to a single entry point, which simplified networking, while reducing exposure and risks for seeing malicious activity.

\subsection{Deployment and Networking}

All services ran as containers managed by Docker Compose. Only Nginx was exposed to the outside network on port~80. The backend, database, and LLM service communicated over an internal Docker network, which reduced the attack surface and allowed services to address each other by name. The frontend interacted with the backend through proxied API routes, and static assets benefited from Nginx caching. This layout kept the public interface relatively secure while allowing internal services to restart or scale independently. In practice, load was very low and no scaling actions were even required.

Runtime configuration was provided through environment variables. Some of the settings included database credentials and the database URL for the backend, the model identifier for the LLM service, and the frontend host URL used for API requests. For our research deployment, however, we used a set of conventional defaults, as we did not need to deploy on multiple environments with different settings.

\subsection{Monitoring and Inspection}

During development and data collection we used a small Python utility, to query selected tables and verify that submissions, feedback, and events were recorded as expected. The script also supported troubleshooting ingestion issues and exporting snapshots of collected data. These checkpoints helped us monitor progress, assess data quality, and decide whether to recruit additional participants when attrition or exclusion, due to malicious behavior, affected our stratified sampling procedure.

%% file: references.bib
@article{wood1976role,
  title={The role of tutoring in problem solving},
  author={Wood, David and Bruner, Jerome S and Ross, Gail},
  journal={Journal of child psychology and psychiatry},
  volume={17},
  number={2},
  pages={89--100},
  year={1976},
  publisher={Blackwell Publishing Ltd Oxford, UK}
}

@article{kalyuga2007expertise,
  title={Expertise reversal effect and its implications for learner-tailored instruction},
  author={Kalyuga, Slava},
  journal={Educational psychology review},
  volume={19},
  number={4},
  pages={509--539},
  year={2007},
  publisher={Springer}
}

@inproceedings{compiler-error-messages-considered-unhelpful,
author = {Becker, Brett A. and Denny, Paul and Pettit, Raymond and Bouchard, Durell and Bouvier, Dennis J. and Harrington, Brian and Kamil, Amir and Karkare, Amey and McDonald, Chris and Osera, Peter-Michael and Pearce, Janice L. and Prather, James},
title = {Compiler Error Messages Considered Unhelpful: The Landscape of Text-Based Programming Error Message Research},
year = {2019},
isbn = {9781450375672},
publisher = {Association for Computing Machinery},
address = {New York, NY, USA},
url = {https://dl.acm.org/doi/10.1145/3344429.3372508},
doi = {10.1145/3344429.3372508},
booktitle = {Proceedings of the Working Group Reports on Innovation and Technology in Computer Science Education},
pages = {177–210},
numpages = {34},
location = {Aberdeen, Scotland Uk},
series = {ITiCSE-WGR '19}
}

@misc{widjojo2023addressingcompilererrorsstack,
  title={Addressing Compiler Errors: Stack Overflow or Large Language Models?}, 
  author={Patricia Widjojo and Christoph Treude},
  year={2023},
  eprint={2307.10793},
  archivePrefix={arXiv},
  primaryClass={cs.SE},
  url={https://arxiv.org/abs/2307.10793}, 
}

@inproceedings{code-debugging-llm-generated-explanations,
author = {Lee, John and Liu, Fengkai and Cai, Tianyuan},
year = {2024},
month = {11},
pages = {1-5},
title = {Code Debugging with LLM-Generated Explanations of Programming Error Messages},
doi = {10.1109/ICEED62316.2024.10923833}
}

@misc{pycee-stack-overflow,
      title={Enhancing Python Compiler Error Messages via Stack Overflow}, 
      author={Emillie Thiselton and Christoph Treude},
      year={2019},
      eprint={1906.11456},
      archivePrefix={arXiv},
      primaryClass={cs.SE},
      url={https://arxiv.org/abs/1906.11456}, 
}

@article{attrition-cs,
author = {Yang, Stephanie and Baird, Miles and O’Rourke, Eleanor and Brennan, Karen and Schneider, Bertrand},
title = {Decoding Debugging Instruction: A Systematic Literature Review of Debugging Interventions},
year = {2024},
issue_date = {December 2024},
publisher = {Association for Computing Machinery},
address = {New York, NY, USA},
volume = {24},
number = {4},
url = {https://dl.acm.org/doi/10.1145/3690652},
doi = {10.1145/3690652},
journal = {ACM Trans. Comput. Educ.},
month = nov,
articleno = {45},
numpages = {44}
}

@article{drop-out-predictions,
author = {Hawlitschek, Anja and Köppen, Veit and Dietrich, André and Zug, Sebastian},
year = {2019},
month = {07},
pages = {},
title = {Drop-out in programming courses – prediction and prevention},
volume = {ahead-of-print},
journal = {Journal of Applied Research in Higher Education},
doi = {10.1108/JARHE-02-2019-0035}
}

@article{what-constitutes-debugging,
author = {Alaboudi, Abdulaziz and LaToza, Thomas D.},
title = {What constitutes debugging? An exploratory study of debugging episodes},
year = {2023},
issue_date = {Sep 2023},
publisher = {Kluwer Academic Publishers},
address = {USA},
volume = {28},
number = {5},
issn = {1382-3256},
url = {https://dl.acm.org/doi/10.1007/s10664-023-10352-5},
doi = {10.1007/s10664-023-10352-5},
journal = {Empirical Softw. Engg.},
month = sep,
numpages = {34}
}

@article{obi2024irritating,
  title={Identifying Factors Contributing to Bad Days for Software Developers: A Mixed Methods Study},
  author={Obi, Ike and Butler, Jenna and Haniyur, Sankeerti and Hassan, Brian and Storey, Margaret-Anne and Murphy, Brendan},
  journal={arXiv preprint arXiv:2410.18379},
  year={2024}
}

@article{debugging-practices,
author = {Perscheid, Michael and Siegmund, Benjamin and Taeumel, Marcel and Hirschfeld, Robert},
title = {Studying the advancement in debugging practice of professional software developers},
year = {2017},
issue_date = {March     2017},
publisher = {Kluwer Academic Publishers},
address = {USA},
volume = {25},
number = {1},
issn = {0963-9314},
url = {https://dl.acm.org/doi/10.1007/s11219-015-9294-2},
doi = {10.1007/s11219-015-9294-2},
journal = {Software Quality Journal},
month = mar,
pages = {83–110},
numpages = {28}
}

@inproceedings{python-multi-paradigm,
author = {Dyer, Robert and Chauhan, Jigyasa},
title = {An exploratory study on the predominant programming paradigms in Python code},
year = {2022},
isbn = {9781450394130},
publisher = {Association for Computing Machinery},
address = {New York, NY, USA},
url = {https://dl.acm.org/doi/10.1145/3540250.3549158},
doi = {10.1145/3540250.3549158},
booktitle = {Proceedings of the 30th ACM Joint European Software Engineering Conference and Symposium on the Foundations of Software Engineering},
pages = {684–695},
numpages = {12},
location = {Singapore, Singapore},
series = {ESEC/FSE 2022}
}

@article{hughes1989functional-matters,
  title={Why functional programming matters},
  author={Hughes, John},
  journal={The computer journal},
  volume={32},
  number={2},
  pages={98--107},
  year={1989},
  publisher={Oxford University Press}
}

@article{conception-evolution-fp,
author = {Hudak, Paul},
title = {Conception, evolution, and application of functional programming languages},
year = {1989},
issue_date = {Sep. 1989},
publisher = {Association for Computing Machinery},
address = {New York, NY, USA},
volume = {21},
number = {3},
issn = {0360-0300},
url = {https://dl.acm.org/doi/10.1145/72551.72554},
doi = {10.1145/72551.72554},
journal = {ACM Comput. Surv.},
month = sep,
pages = {359–411},
numpages = {53}
}

@inproceedings{always-provide-context,
author = {Santos, Eddie Antonio and Prasad, Prajish and Becker, Brett A.},
title = {Always Provide Context: The Effects of Code Context on Programming Error Message Enhancement},
year = {2023},
isbn = {9798400700484},
publisher = {Association for Computing Machinery},
address = {New York, NY, USA},
url = {https://dl.acm.org/doi/10.1145/3576882.3617909},
doi = {10.1145/3576882.3617909},
booktitle = {Proceedings of the ACM Conference on Global Computing Education Vol 1},
pages = {147–153},
numpages = {7},
location = {Hyderabad, India},
series = {CompEd 2023}
}

@inproceedings{not-the-silver-bullet,
author = {Santos, Eddie Antonio and Becker, Brett A.},
title = {Not the Silver Bullet: LLM-enhanced Programming Error Messages are Ineffective in Practice},
year = {2024},
isbn = {9798400711770},
publisher = {Association for Computing Machinery},
address = {New York, NY, USA},
url = {https://dl.acm.org/doi/10.1145/3689535.3689554},
doi = {10.1145/3689535.3689554},
booktitle = {Proceedings of the 2024 Conference on United Kingdom \& Ireland Computing Education Research},
articleno = {5},
numpages = {7},
location = {Manchester, United Kingdom},
series = {UKICER '24}
}

@inproceedings{dcc-help,
author = {Taylor, Andrew and Vassar, Alexandra and Renzella, Jake and Pearce, Hammond},
title = {dcc --help: Transforming the Role of the Compiler by Generating Context-Aware Error Explanations with Large Language Models},
year = {2024},
isbn = {9798400704239},
publisher = {Association for Computing Machinery},
address = {New York, NY, USA},
url = {https://dl.acm.org/doi/10.1145/3626252.3630822},
doi = {10.1145/3626252.3630822},
booktitle = {Proceedings of the 55th ACM Technical Symposium on Computer Science Education V. 1},
pages = {1314–1320},
numpages = {7},
location = {Portland, OR, USA},
series = {SIGCSE 2024}
}

@inproceedings{wang-rct,
author = {Wang, Sierra and Mitchell, John and Piech, Chris},
title = {A Large Scale RCT on Effective Error Messages in CS1},
year = {2024},
isbn = {9798400704239},
publisher = {Association for Computing Machinery},
address = {New York, NY, USA},
url = {https://dl.acm.org/doi/10.1145/3626252.3630764},
doi = {10.1145/3626252.3630764},
booktitle = {Proceedings of the 55th ACM Technical Symposium on Computer Science Education V. 1},
pages = {1395–1401},
numpages = {7},
location = {Portland, OR, USA},
series = {SIGCSE 2024}
}

@inproceedings{toulim-errors,
author = {Barik, Titus and Ford, Denae and Murphy-Hill, Emerson and Parnin, Chris},
title = {How should compilers explain problems to developers?},
year = {2018},
isbn = {9781450355735},
publisher = {Association for Computing Machinery},
address = {New York, NY, USA},
url = {https://dl.acm.org/doi/10.1145/3236024.3236040},
doi = {10.1145/3236024.3236040},
booktitle = {Proceedings of the 2018 26th ACM Joint Meeting on European Software Engineering Conference and Symposium on the Foundations of Software Engineering},
pages = {633–643},
numpages = {11},
location = {Lake Buena Vista, FL, USA},
series = {ESEC/FSE 2018}
}

@article{claim-evidence-reasoning,
  title={Claims, evidence, and reasoning},
  author={McNeill, Katherine L and Martin, Dean M},
  journal={Science and Children},
  volume={48},
  number={8},
  pages={52},
  year={2011},
  publisher={Taylor \& Francis Ltd.}
}

@article{blooms-tax-revision,
  title={A revision of Bloom's taxonomy: An overview},
  author={Krathwohl, David R},
  journal={Theory into practice},
  volume={41},
  number={4},
  pages={212--218},
  year={2002},
  publisher={Taylor \& Francis}
}

@article{armstrong2010bloom,
  title={Bloom’s taxonomy},
  author={Armstrong, Patricia},
  journal={Vanderbilt University Center for Teaching},
  volume={12},
  number={05},
  pages={2023},
  year={2010},
  publisher={Vanderbilt University Nashville, TN, USA}
}

@inproceedings{leinonen-codex,
author = {Leinonen, Juho and Hellas, Arto and Sarsa, Sami and Reeves, Brent and Denny, Paul and Prather, James and Becker, Brett A.},
title = {Using Large Language Models to Enhance Programming Error Messages},
year = {2023},
isbn = {9781450394314},
publisher = {Association for Computing Machinery},
address = {New York, NY, USA},
url = {https://dl.acm.org/doi/10.1145/3545945.3569770},
doi = {10.1145/3545945.3569770},
booktitle = {Proceedings of the 54th ACM Technical Symposium on Computer Science Education V. 1},
pages = {563–569},
numpages = {7},
location = {Toronto ON, Canada},
series = {SIGCSE 2023}
}

@misc{salmon2025debuggingerrormessagesllm,
      title={Debugging Without Error Messages: How LLM Prompting Strategy Affects Programming Error Explanation Effectiveness}, 
      author={Audrey Salmon and Katie Hammer and Eddie Antonio Santos and Brett A. Becker},
      year={2025},
      eprint={2501.05706},
      archivePrefix={arXiv},
      primaryClass={cs.SE},
      url={https://arxiv.org/abs/2501.05706}, 
}

@article{charlesEffectAutomatedError2023,
  title = {The {{Effect}} of {{Automated Error Message Feedback}} on {{Undergraduate Physics Students Learning Python}}: {{Reducing Anxiety}} and {{Building Confidence}}},
  author = {Charles, Tessa and Gwilliam, Carl},
  year = {2023},
  month = aug,
  journal = {Journal for STEM Education Research},
  volume = {6},
  number = {2},
  pages = {326--357},
  issn = {2520-8713},
  doi = {10.1007/s41979-022-00084-4}
}

@incollection{sweller2011cognitive,
  title={Cognitive load theory},
  author={Sweller, John},
  booktitle={Psychology of learning and motivation},
  volume={55},
  pages={37--76},
  year={2011},
  publisher={Elsevier}
}
